\RequirePackage{fix-cm}

\documentclass[smallextended]{svjour3}       
\smartqed  
\usepackage[square,numbers]{natbib}
\usepackage{graphicx}

 \journalname{my journal}
\begin{document}
\bibliographystyle{plainnat}
\title{Critique of Quantum Optical Experimental Refutations of Bohr's Principle of Complementarity, of the Wootters-Zurek Principle of Complementarity, and of the Particle-Wave Duality Relation
}

\titlerunning{Critique of  Experimental Refutations  of  Complementarity}        

\author{P. N. Kaloyerou    
}


\institute{P. N. Kaloyerou \at
              The University of Zambia\\ School of Natural Sciences\\              Department of Physics\\Lusaka 10101\\ Zambia\\ 
              Tel.: +260-977-566-031\\
              \email{pan.kaloyerou@wolfson.ox.ac.uk.}           
           \and
and Wolfson College, Linton Road, Oxford OX2 6UD, UK. 
}

\date{Received: date / Accepted: date}

\maketitle

\begin{abstract}
I argue that quantum optical experiments that purport to refute Bohr's principle of complementarity (BPC) fail in their aim. Some of these experiments try to refute complementarity by refuting the so called particle-wave duality relations, which evolved from the Wootters-Zurek reformulation of BPC (WZPC). I therefore consider it important for my forgoing arguments to first recall the essential tenets of BPC, and to clearly separate BPC from WZPC, which I will argue is a direct contradiction of BPC. This leads to a need to consider the meaning of particle-wave duality relations and to question their fundamental status. I further argue (albeit, in opposition to BPC) that particle and wave complementary concepts are on a different footing than other pairs of complementary concepts. 
\end{abstract}

\section{Introduction}
In recent years there has been a proliferation of experimental tests of Bohr's principle of complementarity (henceforth BPC) \cite{BR28, Bohr34, BohrPC, Schilpp, JAM74}. In reality, such tests nearly always concern particle and wave complementary concepts. Many of the tests are optical tests. Representative and important examples of such optical tests are in references \cite{GHA91, BGGP04, MO92, AFSHAR04,  G86}. Rauch et al's use of a single silicon crystal, suitably cut, as a neutron interferometer  \cite{RTB74} allowed many experimental tests of aspects of particle-wave duality using neutrons (see for example \cite{RW, RS84,GY88, KB1992, BCSK92, BRT86}). More recently, experiments on the particle-wave aspects of atoms have been made possible with the development of the atom interferometer \cite{BB1996, DNR1998NAT, LI2001,DNR98PR,  B2001, SZ97, MSK2008}.  Further references  of tests of complementarity may be found in Ghose's book \cite{Ghose1999}. 

Many of the tests of complementarity give results which authors claim confirm complementarity. These include the neutron interferometry tests, the atom interferometer tests and many of the optical tests. A notable optical test which claims to confirm complementarity is that of Grangier, Roger and Aspect (GRA) \cite{G86}. The GRA experiment is notable because it was perhaps the first such experiment to use  a gating system to produce genuine single photon states (Fock states) using photons produced  by atomic cascades. However, GRA fail in their objective to confirm complementarity because, although their results are consistent with complementarity, their results can just as well be explained by the Bohm-de Broglie causal interpretation  \cite{B52, DEBR60}, hereafter referred to  simply as the causal interpretation,  or its extension to the electromagnetic field \cite{K06}. All of the experiments which claim to confirm complementarity can be described in terms of the causal interpretation, or its extension to boson fields \cite{K06, K85, BHK87, K94, PNK2005}. Any experiment consistent with two or more alternative interpretations or theories cannot be regarded as proof of one them.

Before proceeding to my main aims, I review other experimental tests of BPC, and also very interesting related experiments  which push the boundaries of complementarity (and which present challenges to alternative interpretations of the quantum theory): The Wheeler delayed choice experiment, and quantum erasure experiments.

A novel  test of complementarity uses electrons following two-paths in an Aharanov-Bohm ring interferometer  \cite{QDOT1, QDOT2}. After the electrons traverse the two paths they are recombined to produce interference effects. A quantum dot embedded in one path and coupled to a quantum-point-contact charge detector is used for path detection. Their results are consistent with BPC.

The interference of atoms raises the further question of what happens to their internal structure in an interference experiment. The same question applies, of course, to fundamental particles with their proposed quark structures, but the question seems to be particularly poignant with atoms. Further, the internal structure can be used to mark the path. As expected, when the path is identified interference is lost in agreement with BPC \cite{BB1996, DNR1998NAT, LI2001,  B2001, SZ97}. Intermediate experiments, where the path is only partially determined resulting in reduced visibility interference, have also been performed with results again consistent with complementarity \cite{DNR98PR, MSK2008}. A particular question authors focused on is how complementarity is enforced. In many tests, complementarity is enforced by the position-momentum uncertainty relations. But, in atom interferometer experiments, authors such as D\"{u}rr et al \cite{DNR1998NAT} and Li \cite{LI2001} have demonstrated very clearly that complementarity is enforced by entanglement, rather than by the uncertainty relations. Specifically, an entangled state between the internal degrees of freedom and the possible paths is formed such that interference terms disappear once the path is determined.  If we adhere strictly to BPC, then the question raised above concerning what happens to the internal structure when atoms interfere cannot be asked. This is so for two reason: First, in BPC, physical reality cannot be attributed to classical concepts. Secondly, in BPC, an experiment must viewed as an undivided whole, not further analyzable. However, very few workers, including the present author, adhere to this strict view and tend to think of an underlying physical reality. So, although such atom experiments with markers are consistent with BPC, they surely question the plausibility of BPC. This point is highlighted  with the description of atom interferometry experiments according to the causal interpretation. Here, when atoms form an interference pattern, the atoms remain localized and retain their internal structure throughout, Interference is produced by the $R$ and $S$-fields associated with quantum state of the atoms (The form of the $R$ and $S$-fields depends on the boundary conditions).

Wheeler added another dimension to the experimental tests with his delayed-choice experiment. The decision of whether to measure path or interference is left until the last instant. This leads either to the paradoxical conclusion that history is changed at the time of measurement, or to Wheeler's preferred interpretation, that history is created at the instant of measurement. Thus Wheeler writes,`` No phenomenon is a phenomenon until it is an observed phenomenon'' \cite{WHR78} pp. 14. He adds that ``Registering equipment operating in the here and now has an undeniable part in bringing about that which appears to have happened'' \cite{WHR83} pp. 194. Wheeler concludes, ``There is a strange sense in which this is a `participatory universe'.''  \cite{WHR83} pp. 194. That this experiment is consistent with BPC is guaranteed by a central tenet of BPC, namely, classical concepts in the quantum theory are  abstractions to aid thought and to communicate the results of experiment, but cannot be attributed physical reality. Wheeler \cite{WHR78, WHR83} appears not to adopt the latter tenet, but instead follows Heisenberg \cite{HEIS83}, and, in some sense, attributes reality to complementary concepts (and to the wave function), hence the paradoxical conclusion. We note that Bohr anticipated delayed choice experiments, he writes ``\ldots it obviously can make no difference as regards observable effects obtainable by a definite experimental arrangement, whether our plans of constructing or handling the instruments are fixed beforehand or whether we prefer to postpone the completion of our planning until a later moment when the particle is already on its  way from one instrument to another.'' \cite{Schilpp} pp. 230. Again, both the causal interpretation \cite{BDH85} and its extension to boson fields \cite{PNK2005} can explain the Wheeler delayed-choice experiment in a causal, non-paradoxical way.

A further push of conceptual boundaries occurred with the introduction of quantum erasure experiments \cite{SD82, ZWZM91, SEW91, KSC92, HKWZ95, ESW99, KKSS2000}. Perhaps the best example is the quantum eraser experiment of Kim et al \cite{KKSS2000}. I base my few comments on this experiment.  The two slits of a two-slit experiment are replaced by two sources producing entangled photon pairs. One photon of an entangled photon pair travels backwards  carrying the path information of its forward moving partner photon,  and is directed to one of four detectors by a system of mirrors and  beam splitters.  For a photon detected in either of the two outer detectors, the source from which it was produced is revealed so the path of its entangled partner is determined with certainty, i.e., photons detected in either of the two outer detectors retain path information.  On the other hand, a photon detected in either of the two inner detectors passes through an extra beam splitter, so it may have come from either source, i.e., its path information is erased by the mixing of paths at the beam splitter. Consequently, the path of its entangled photon partner remains completely undetermined. The other photon of the entangled pair travels forwards  toward an interference detector, where interference fringes can be observed. The detection of the entangled photon pairs is correlated. It was found that for the backward photons which retain path information, their entangled forward partners did not form interference fringes, while for those backward photons for which path information was erased, their entangled  forward partners did form interference fringes (click by click). What makes the experiment so paradoxical is that the forward photons reach the interference detector long before their  backward entangled partners reach a beam splitter. Therefore, interference occurs (or not) long before, the path information is erased (or not). \textit{This implies that a measurement in the present affects a measurement in the past}. In the authors opinion, this is  impossible and is not implied by quantum theory. The reason is that nonlocality is ``once-only'', in other words, the entanglement between the two photons is broken upon the first detection, which is the detection of the forward photon at the interference detector.  If nonlocality was not like this, and  persisted after the first detection it would be possible to set up causal paradoxes and  faster than light signaling leading to  the possibility of experimental contradiction with relativity. Thus, once a photon reaches the interference detector, its entangled  connection with its backward traveling partner is broken. Whatever happens to this  backward photon, thereafter, cannot affect the forward detected photon. To justify this view further, the experimental results have to be calculated based on the quantum feature  of broken entanglement (or once-only-nonlocality).

Many of the experimental tests of complementarity do not deal directly with BPC, but deal instead with the particle-wave duality relation  which has evolved from the Wootters-Zurek principle of complementarity (WZPC) \cite{WZ79}. In such tests authors equate the duality relation with complementarity, and claim that complementarity is refuted because the duality relation is refuted.  WZPC has given rise to the so called ``intermediate experiments'', in which the path is only partially determined, resulting in reduced visibility of the interference fringes\footnote{See reference \cite{BART80} for an early proposal for realisable intermediate experiments.}. This, in turn, has  led to the development of various forms of the particle-wave duality relation. I will argue in Sect. \ref{WZPC1} that WZPC is a contradiction of BPC, and that WZPC is only meaningfully interpreted with reference to an ontology, but in this case WZPC loses its fundamental significance. Similarly, care is also needed in interpreting  the duality relation if contradiction is to be avoided. Referred to the same experiment, the duality relation amounts to a contradiction of definitions of the two mutually exclusive classical concepts of wave and particle. However, the duality relation can be given a meaning consistent with BPC by interpreting it with respect to two mutually exclusive experimental arrangements, as suggested by  Jaeger, Shimony and  Vaidman (JSV) \cite{JSV95}. Alternatively, as with WZPC, the duality relation can be given meaning with reference to an ontology. However, with either of the latter interpretations the duality relation loses its fundamental conceptual significance. From the mathematical perspective the duality relation cannot be ignored, and has been derived from quantum theory by JSV \cite{JSV95}, with a particularly rigorous derivation given by Englert \cite{ENG96}.  Though the particle-wave duality relation does not have the conceptual significance normally attributed to it (at least in the authors opinion), it is clear that it does have mathematical significance. Therefore, a refutation of the duality relation would amount also to a refutation of the quantum theory. Even though a number of experimental tests of BPC are discussed in terms of the duality relation, all such tests are ultimately direct tests of BPC. In this article, we will consider all experiments with reference to BPC, but emphasise that the same reasoning that saves BPC also saves the duality relation.

My main aim, as I have said, is to show that the experiments which purport to refute complementarity fail, irrespective of whether the results are considered directly in terms of  BPC or indirectly in terms of the particle-wave duality relations. But from what I have said above, there are crucial preliminaries that need to be clarified. 

First, there is a need to  emphasize an aspect of BPC that has been universally overlooked, and that Bohr would never agree with, namely, that particle and wave complementary concepts are fundamentally different from other pairs of complementary concepts

Second, we must clearly separate BPC and WZPC.  To do this we first remind ourselves of the main tenets of BPC, and justify these tenets with some quotes from the great man himself. Though I am here defending BPC from experimental refutation, I will make some critical remarks against BPC on general theoretical grounds. Following this, I consider WZPC and provide reasons why it must be viewed as a separate principle from BPC, indeed, why it is a direct contradiction of BPC.  Note that the term complementarity has been extended beyond the classical concepts envisaged by Bohr. An example is the complementarity between single photon and two-photon interference \cite{JHS93, JSV95}. Therefore, I will use the abbreviation ``BPC'' to refer to  Bohr's original strict version, and use the term ``complementarity'' in contexts where a  more general  usage is appropriate. 

Third, I consider the particle-wave duality relations  \cite{GY88,  JSV95, ENG96, SM89, ZWM91,  M91,  LHGZ2012}. 

For my main aim, I select the following experiments for consideration: The  1991 Ghose, Home and Agarwal experiment (GHA) \cite{GHA91},  later performed by Mizobuchi and Ohtak\'{e} \cite{MO92}; the Brida, Genovese,  Gramegna, and  Predazzi  experiment (BGGP) \cite{BGGP04}; and the Afshar experiment  \cite{AFSHAR04}. The Afshar experiment, in particular, has received considerable attention, so I will consider this experiment in greater detail.  

In these experimental refutations, one complementary concept is actually measured (i.e. defined by the experimental arrangement as required by BPC), while the other complementary property is artificially introduced through an unjustified assumption or assumptions. In such experiments the experimenters include an intermediate process, which in classical theory must  be described by the  classical concept opposite to the classical concept consistent with the final actual measurement. But since this concept is not actually measured, it is never defined in the way demanded by BPC. Thus, BPC (and even the duality relations) are not even nearly challenged by these experiments. Important to my defense of BPC is the consideration of what constitutes a measurement and what constitutes an inference, based on an assumption or assumptions, drawn from the experimental arrangement. I will address this question in Sect. \ref{WISAM}, together with a moderatley brief overview of quantum mechanical measurement theory, including a brief outline of my preferred solution of the measurement problem.

I begin by considering particle and wave complementary concepts followed by BPC, WZPC, particle-wave duality relations, and then, in separate sections, the experiments of AHG, BGGP, and Afshar.  

\section{Particle-Wave Complementary Concepts}
\label{PWCC}
Following many authors before me, I begin with Feynman's famous quote concerning particle-wave duality, \cite{F64} vol. III, pp. 1-1, ``In reality, it contains the only mystery". I have emphasized this quote because it relates to the first crucial point I wish to emphasize, namely, that particle and wave complementary concepts are fundamentally different from other complementary concepts, such as position and momentum, which classically, are canonically conjugate dynamical variables (CCDV). Complementary concepts which are CCDV are identified with definite elements of the mathematical formalism of quantum mechanics, namely, linear hermitian operators, and satisfy uncertainty relations rigorously derivable from the quantum theory. Crucially, CCDV are {\bf NOT} mutually exclusive classical concepts. Classically, such variables have simultaneously well defined values at all times which precisely define the state of a classical system. On the other hand, the concepts of wave (field) and particle are mutually exclusive classical concepts. It is simply a contradiction of definitions to describe a single object  as a wave and a particle. Moreover, Bohr never identified  particle and wave concepts with mathematical elements of the quantum formalism.  Bohr, of course, never accepted such a differentiation of complementary concepts, but instead sought a unified view with all pairs of complementary concepts on an equal footing. This is very likely, why, Bohr never viewed complementarity as identical with Heisenberg's uncertainty relations. Jammer emphasized this point in his book \cite{JAM74} and writes,  ``That complementarity and Heisenberg-indeterminacy are certainly not synonymous follows from the simple fact that the latter... is an immediate mathematical consequence of the {\it formalism} of quantum mechanics or, more precisely, of the Dirac-Jordan transformation theory, whereas complementarity is an extraneous {\it interpretative} addition to it"  \cite{JAM74} pp. 61.  The importance of this separation is that it clarifies subsequent discussion, and it also removes the perceived need for a ``missing'' particle-wave uncertainty relation. It is difficult to see  how a conceptually non-contradictory interpretation of such an uncertainty relation could be given. Note that the particle-wave duality relation differs from uncertainty relations, as has been emphasized by Englert \cite{ENG96}. 

This distinction has important conceptual consequences. CCDV complementary concepts, which classically are  simultaneously measurable and conceivable, are also simultaneously conceivable in the quantum theory. The uncertainty relations place a limit on their simultaneous measureability, but not on their simultaneous conceiveability. Thus, it is easy to picture a measurement of, say, position disturbing a previously known value of momentum so the that the new value of momentum is not known, but this does not prevent us in the quantum theory from simultaneously picturing a particle with a well defined position (with a value known by the measurement) and a well defined momentum (with an unknown value because of the disturbance by the position measurement). But, with the mutually exclusive concepts of particle and wave, picturing a single object as a wave and a particle is not possible. This is just as true in quantum theory as it is in classical theory. A single object is simply not simultaneously picturable as  a wave and a particle. This impossibility is the root behind the particle-wave duality paradox, and no wonder coined by Feynman ``the only mystery''. In my view, this paradox or ``mystery'' is only resolved by an ontological interpretation of the quantum theory, such as the causal interpretation. 

\section{Bohr's Principle of Complementarity\label{BPC28}}
\label{BPC27}
Not withstanding the countless articles relating to complementarity, I believe that it is still worthwhile to recall the main tenets of Bohr's (original) principle of complementarity.  As is well known, Bohr's first presentation of complementarity in a fairly complete form was in his 
1927 Como lecture \cite{BR28}. Further presentations of BPC can be found in references \cite{Bohr34, BohrPC, Schilpp, JAM74}. 

The four core tenets of Bohr's Principle of Complementarity are as follows:
\begin{enumerate}
\item[(T1)] The concept of a precisely definable classical state must be given up (because of the quantum postulate), and a separation of subject (experimental apparatus) and object (quantum system) is impossible. An experiment must therefore be viewed as an unanalyzable whole.
\item[(T2)] A single picture is not sufficient to exhaust the description of a quantum system. Rather pairs of complementary concepts are needed. Such complementarity concepts (for example, wave and particle concepts) can only be used in  mutually exclusive experimental arrangements. Indeed, it is the experimental arrangement which defines the concept to be used.
\item[(T3)] Classical concepts are abstractions to aid thought and to communicate the results of experiment, but cannot be attributed physical reality.
\item[(T4)] A description of physical processes  that underlie experiment in terms of a single well defined model is impossible, i.e., a description
of  underlying physical reality in terms of a single-well defined model is impossible.
\end{enumerate}
T1 stems from Bohr's quantum postulate, which  states that a quantum system interacts with its environment (e.g. measuring apparatus) through the exchange of a quantum of action that is indivisible, uncontrollable, and unanalyzable. As a consequence, the concept of a classical state defined by a complete set of well defined dynamical variables has to be given up. This leads to the conclusion that the apparatus and quantum system must be viewed as an unanalysable whole. Bohr emphasized that it is the experimental arrangement which defines the property being measured. Possibly, T2 was motivated by the photoelectric and Compton effects, which were interpreted as  indicating a particle  behaviour of light, whilst, hitherto, classical experiments revealed a wave nature of light, and by the Davisson-Germer experiments in which electron beams produced  interference patterns, hence revealing a wave nature of electrons, whlist earlier experiments indicated electrons had a particle nature.
It seems reasonable to conjecture that T3 was motivated by the mutually exclusive nature of wave and particle concepts. Perhaps, Bohr felt that the application of wave and particle concepts to  the same physical object is a contradiction of definitions, even if, as he stated, they cannot both be applied in the same  experiment. It appears, especially with regard to particle and wave complementary concepts, that Bohr's aim was to provide a framework for the non-contradictory use of classical language/concepts. The latter point is indicated in the following quotes from Bohr, ``...in dealing  with the task of bringing order into an entirely new field of experience, we could hardly trust in any accustomed principles, however broad, apart from the demand of avoiding logical inconsistencies..." \cite{Schilpp} pp. 228, and ``...we cannot seek a physical explanation in the customary sense, but all we can demand in a new field of experience is the removal of any apparent contradiction.'' \cite{BohrPC} pp. 90. 

I finish this brief summary  with some more quotes from Bohr's writings:
\begin{enumerate} 
\item[(i)] In the following quote, relating to T1 and indicating  T4, Bohr, referring to the quantum theory, writes, ``its essence may be expressed in the so-called quantum postulate, which attributes to any atomic process an essential discontinuity, or rather individuality, completely foreign to the classical theories and symbolized by Planck's quantum of action. This postulate implies a renunciation as regards the causal space-time co-ordination of atomic processes. Indeed, our usual description of physical phenomena is based entirely on the idea that the  phenomena concerned may be observed without disturbing them appreciably. \ldots Accordingly, an independent reality in the ordinary physical sense can neither be ascribed to the phenomena nor the agencies of observation.'' \cite{BR28} (b) pp. 580.
\item[(ii)] With regard to T2, we recall the following quote from Bohr:  ``\ldots however far the phenomena transcend the scope of classical physical explanation the account of all evidence must be expressed in classical terms"\ldots ``this crucial point\ldots implies the impossibility of any sharp separation between the behaviour of atomic objects and the interaction with the measuring instruments which serve to define the conditions under which the phenomena appear. \ldots Consequently, evidence obtained under different experimental conditions cannot be comprehended within a single picture but must be regarded as complementary in the sense that only the totality of the phenomena exhausts the possible information about the objects. Under these circumstances an essential element of ambiguity is involved in ascribing conventional physical attributes to atomic objects, as is at once evident in the dilemma regarding the corpuscular and wave properties of electrons and photons, where we have to make do with contrasting pictures, each referring to an essential aspect of empirical evidence."  \cite{Schilpp} pp.  209-210. Regarding the measurement of complementary concepts Bohr writes, ``As repeatedly stressed, the principal point here is that such measurements demand mutually exclusive experimental arrangements." \cite{Schilpp}  pp. 233 and \cite{BohrPC} pp. 60. \\ \mbox{}\\
\item[(iii)] Concerning the reality of particle or wave concepts (referred to in T3) he writes, ``radiation in free space as well as isolated material particles are abstractions, their properties on the quantum theory being definable and observable only through their interaction with other systems. Nevertheless, these abstractions are \ldots indispensable for a description of experience in connexion with our ordinary space-time view." \cite{Bohr34} pp. 57.
\end{enumerate}

Proponents of  BPC may well consider my reduction of BPC to four tenets as anathema. But, I have taken the step  to reduce BPC to its essential core content, as expressed in my for tenets,  deliberately for reasons I will try to explain. I will also try to justify my choice of tenets, since it is unlikely that all authors would agree with my choice, especially T4. It is well known that there is no consensus as to the meaning of BPC. It is sometimes stated, perhaps flippantly, that there are as many interpretations of complementarity as there are physicists. Though there may be many varied interpretations of BPC, it does not necessarily mean that all the interpretations, and developments of BPC (especially more modern versions) are fully consistent with BPC. For example,  it is not at all clear that even Bohr's contemporaries (e.g. Heisenberg, Von Neumann, Wigner) interpreted BPC in a way that was fully consistent with Bohr's own expression of his principle. Even Bohr, himself, never reached a final version of complementarity that he was satisfied with. I  have thus come to the view that attempts to state BPC in an over precise way leads to less clarity, and perhaps, leads away from Bohr's original exposition, if not to a contradiction of Bohr's view. But BPC does have an essence, a bare-bones meaning if you like. It is my view that it is in its bare-bones form that BPC takes on its most useful form as a guide to working physicists in discussing experiments. I believe that my four tenets, though intuitively stated and seemingly imprecise, do capture the essence of BPC, and are indeed consistent with Bohr's original writings (though not necessarily with other versions of complementarity, especially more modern versions). Though I am not a proponent of BPC (so it may seem ironic that I am defending BPC in this article, but, it is just that, I believe that if BPC is to be rejected, it should be rejected for the right reasons),  I think that BPC was an important stepping stone in the understanding of quantum mechanics for two reasons: (1) In his principle, Bohr recognised the central new feature of quantum mechanics, namely, the property of  "wholeness" or "interconnectness" that is inherent  in quantum theory. This recognition, in my view, was an act of genius and a vital element in understanding quantum theory, and (2) BPC provides a framework for the non-contradictory use of classical concepts. Here, a clear distinction must be made between classical concepts and classical theory. Bohr insisted on the use of classical concepts to describe experiment, not classical theory. The use of the latter would make no sense, since classical theory failed to describe many experiments, a failure that inspired the introduction of the quantum theory (as well as special and general relativity) in the first place.

I suspect that tenet T4 may be subject to the most disagreement. Yet, it is the inescapable consequence of the other three tenets. If experiments must be viewed as unanalysable wholes (T1), if pairs of complementary classical concepts are needed to exhaust the description of nature (T2), and if physical reality cannot be attached to classical concepts (T3), then it surely follows that a description of underlying physical reality in terms of a single well defined model is denied (tenet T4). It might be argued that simply attributing an objective reality to the wave function is an argument against T4 being regarded as a part of BPC. First, simply attributing reality to the wave function without further interpretational elements doesn't constitute a model of underlying physical reality. For example, interpreting the wave function as an objectively existing probability wave has no explanatory value. In this view, what causes the spot on the photographic plate? What is an atom?  Probability generally refers to an underlying ontology, but does not of itself provide a model of reality. But second, and more relevant, is that attributing an objective reality to the wave function as a model of reality directly contradicts Bohr's insistence that the world must be described in terms of classical concepts, i.e., the assertion contradicts T2.  Further, as an adherent of the causal interpretation, I certainly attribute reality  to the wave function. But reality of the wave function alone does not lead to the causal interpretation and further interpretational elements are needed. But, the existence of the causal interpretation doesn't constitute an argument against T4 being a part of BPC, but instead, is an argument against BPC in its entirety. Therefore, attributing reality to the wave function cannot be used as an argument to exclude T4 as a part of BPC, instead it can be used to argue against BPC in its entirety.

I have not included the so called ``classical cut'' between the classical and quantum domains as part of BPC, simply because it never featured in Bohr's writings, i.e., it never formed a part of BPC as presented by Bohr. Rather, the idea of the "cut" was introduced and emphasised by Bohr's followers, especially   Heisenberg \cite{H52a, H52b,  CS2015}. Indeed, the "cut" was a point of private disagreement between Bohr and Heisenberg. Though Bohr never publicly rejected the idea of the "cut", neither did he endorse it \cite{CS2015} pp. 6.

The idea of a ``cut'' seems to be contrary to the BPC assertion (T1) that an experiment must be viewed as a undivided whole, not further analysable. This implies that Bohr viewed the classical and quantum worlds as intricately  connected in such a way that they cannot be separated. Whatever processes occur at the quantum level underlying experiment is completely beyond our conceptual gaze according to BPC, as is how quantum processes conspire to form the classical world. That the notion of a cut did not form a part of BPC is further enforced by the suggestion that Bohr viewed the quantum theory as no more than a probability calculus, and did not attribute objective reality to the wave function\footnote{This view was expressed by Prof. D. Bohm in private discussions during the years 1980 to 1985. Though I have never found a categoric statement in Bohr's writings to support this view, there are indications in his writings pointing to this view, for example, Bohr writes, ``...the appropriate physical interpretation of the symbolic quantum-mechanical formalism amounts only to predictions, of determinate or statistical character, pertaining to individual phenomena appearing under conditions defined by classical physical concepts.'' \cite{Schilpp} pp. 238. \label{FF2}}. In this case, there is no need to introduce a idea of a cut, since only the macroscopic phenomena  enter in the description of nature, phenomena which occur with probabilities calculated from the wave function.

As an alternative to my reduction of BPC to four  bare bones tenets, I might point the reader to a recent article by Camilleri and Schlosshauer \cite{CS2015}, in which the authors   present a much more detailed exposition of BPC which, like the present author, is based on Bohr's original writings. This article also contains a fairly comprehensive list of references relating to BPC.

Since we are concerned with experimental tests of BPC, we need to  consider operational consequences of BPC as expressed in the above four tents. Tenet T1 has been tested and confirmed a countless number of times by everyday quantum mechanical experiments. For example, it is straightforward to construct a Stern-Gerlach experiment to change the spin state of a beam  of electrons, and to subsequently check that the change,  except for the special case of eigenstates, is uncontrollable, unpredictable, and unanalysable in conformity with T1. The operational consequence of tenet T2 is the purpose of this article to discuss. If an experiment can be found in which complementary concepts are measured in the same experiment, this would contradict tenet T2 and hence BPC. No such experiment has yet been conducted. The particular experimental tests that we will consider concern attempts to measure (detect) particle and wave behaviour in the same experiment. If the latter were to be achieved, this again would  contradict tenet T2, and BPC.

Tenet T3 is a prescription for describing experiments in a non-contradictory way. T3 is thus an interpretational element. A test of T3 is therefore a test between different interpretations of the quantum theory. However, for interpretations that are consistent with the quantum theory, any experiment whose results are consistent with the quantum theory cannot distinguish between the different interpretations.  Different interpretations could, however, suggest newer experiments. If the results of these newer experiments differed from the quantum theory, these results may differentiate between the different interpretations. We conclude that an experimental test of T3 amounts to an experimental test of quantum theory itself.

The correctness of tenet  T4 and hence BPC in its entirety, since T4 is an inescapable consequence of T1, T2 and T3, is directly challenged theoretically by the existence of the causal interpretation, as we mentioned above. But again, an experimental test would amount to an experiment to distinguish between interpretations and similar considerations as in the above paragraph apply. Therefore, we again conclude that an experimental test of T4 amounts to an experimental test of quantum theory itself.

\section{Critique of Bohr's Principle of Complementarity}
As I have stated, my main aim is to strongly defend BPC against experimental refutation. I do this because I believe that if BPC is to be rejected it should be rejected for the right reasons. I think that there are strong theoretical and conceptual grounds for rejecting complementarity. An experimental refutation would require two actual measurements, each consistent with the opposite complementary concept. I strongly suspect that if such an experiment is ever achieved, it would also refute the quantum theory. One criticism of BPC concerns the fact, discussed in some detail in Sect. \ref{PWCC}, that particle and wave complementary concepts are fundamentally different to other complementary concepts. This renders the unified view sought by Bohr  somewhat strained. Further, describing the same object by two mutually exclusive concepts, albeit in mutually exclusive experiments, is of itself inherently contradictory. No wonder, to achieve consistency,  Bohr did not attribute physical reality to complementary concepts. Such a description hardly constitutes an adequate description of nature. The causal interpretation also contains both a wave and particle aspect but in a fundamentally different way: electrons, protons, neutrons etc, are particles in every experiment, while the wave-aspect is always associated with $R(x,t)$ and $S(x,t)$  fields.  

A more serious objection concerns my argument, also discussed in Sect. \ref{PWCC}, that BPC is not a direct interpretation of the  mathematical  formalism of the quantum theory.  As I mentioned, though CCDV are identified with linear hermitian operators, particle and wave concepts in BPC are never identified with any element of the mathematical formalism.  It might be thought that the recently (relatively) introduced  particle-wave duality relation  answers the latter failing, but this view is subject to criticism, as we argue in Sect. \ref{SECPWDR}.

A final serious criticism of BPC is the requirement that a description of underlying physical reality is impossible (tenet T4). This requirement is essential for the consistency of BPC. This  is an extremely high price to pay for consistency. Fortunately, the existence of the  causal interpretation based on which computer models of underlying physical reality can be produced \cite{DEWD,QPPLOTS, K89} not only shows that we are not forced to accept this extreme position, but also that it is wrong (though I note that recently, the reality of the trajectories in the causal interpretation has been questioned \cite{ENGSSW92}). As mentioned in Sect.  \ref{BPC28}, not all authors would agree with tenet T4. For an alternative discussion of the reality of trajectories and BPC we refer the reader to an interesting recent article by Drezet \cite{DRZ14} \footnote{In his article Drezet  discusses the reality of trajectories and BPC in the context of arguing that the recently introduced Pusey, Barret and Rudolf (PBR) Theorem, concerned with ontic and epistemic hidden variable theories, does not preclude the causal interpretation (sometimes also referred to as Bohmian mechanics).}.

\section{The Wootters-Zurek Principle of Complementarity\label{WZPC1}}
In their 1979 article \cite{WZ79} WZ analysed Einstein's two-slit experiment from the perspective of partial particle and partial wave information using Shannon's measure of information, defined as
\[
H=-\sum_{i=1}^{N}p_i \ln (p_i),
\]  
where $H$ is a positive number giving the ``information we lack'', $N$ is the number of possible states of the system, and $p_i$ is the probability of the system being in  state $i$. Using this definition, WZ developed a reciprocity measure of how much wave and particle information  can be obtained from the same experiment.  Their analysis led WZ to the following reformulation of complementarity: {\it The sharpness of the interference pattern can be regarded as a measure of how wave-like the light is, and the amount of information we have obtained about the photon's trajectories can be regarded as a measure of how particle-like it is}. I shall call this reformulation WZPC.

Mathematically, the WZ results are very interesting, but as I argued in my 1992 article  \cite{K92}, WZPC embodies everything that Bohr warned against. We recall the following emphasis from Bohr, ``The argument is simply that by the word ``experiment'' we refer to a situation where we can tell others what we have done and what we have learned and that, therefore, the account of the experimental arrangement and of the results of the observations must be expressed in unambiguous language with suitable application of the terminology of classical physics."\cite{Schilpp} pp. 209. WZPC hardly fulfills this condition. 

We expect of a physical concept (together with its mathematical representation) to have an explanatory (and predictive) function. In classical physics the concept of wave and its mathematical representation explains interference. The visibility of the fringes has nothing to do with the degree to which the wave concept applies, but rather, fringe visibility is a measure of the degree of coherence of the particular wave profile. Even the tiniest level of visibility of interference fringes requires a 100\% wave model; not 50\% or 20\%. What possible meaning can be given to the partial application of the wave concept. I conclude,  that so called intermediate experiments in which partial path information is obtained at the expense of visibility (which WZ related to the degree of wave knowledge) are {\bf NOT} intermediate at all. Even where the tiniest fringe visibility is observed, BPC clearly demands a 100\% wave model. Attributing a particle concept to such experiments is arbitrary and artificial and certainly in contradiction with BPC. For example, in a two slit experiment with one slit half the size of the other, the fringe visibility would certainly be reduced. Classically, we say that twice as much wave energy passed through the bigger slit. WZPC would attribute a probability of 0.33 to the path through the smaller slit and a probability 0.67 to the path through the larger slit, then substitute these values into their formula  to determine the path information. But this allocation is artificial, since the experiment does not define the particle concept in the way specified by BPC. 

For these  reasons I conclude that WZPC is a contradiction of BPC. This conclusion does not mean that the WZ analysis based on information theory is not an interesting and valuable analysis; our objection is only in the interpretation of the analysis, and in the claim that it is a reformulation of BPC. Henceforth, BPC and WZPC will be considered to be two entirely separate principles.

Perhaps ironically, WZPC can be consistently interpreted using the nonrelativistic causal interpretation \cite{B52, DEBR60} (but not its extension to boson fields\footnote{Photons are bosons and are governed by the second quantized Maxwell equations. In the causal interpretation of boson fields, fundamental entities are purely fields; there are no boson particles. In this case WZPC cannot be given meaning based on this ontology\label{CIBF}.}). In other words, WZPC can be given a conceptually consistent interpretation with reference to an appropriate ontology. This is not a surprise, since classical probability and classical information theory refer to an underlying ontology. In the causal interpretation electrons, protons etc (but not photons, for the reasons given in footnote \ref{CIBF}) are particles with associated guiding fields (the $R$- and $S$-fields\footnote{The $R$ and $S$-fields defined by $\psi=R\exp(iS/\hbar)$ are not independent of each other. Both play an equal part in determining a particles motion, the $R$-field through the quantum potential $Q= -\hbar^2/(2m)\nabla^2 R/R$, and the $S$-field through the guidance formula $v=\nabla S/m$.}). This model contains a 100\% particle concept  and a 100\% wave concept. In the two-slit experiment, an electron, say, passes through only one slit, and is then guided by  the $R$- and $S$-fields  to a bright fringe. In intermediate experiments it now  becomes meaningful (and non-artificial) to attribute a probability to each possible  path. Therefore, a particle is a 100\% a particle, but its path in a particular experiment may not be determined with 100\%  certainty\footnote{Actually, in computer models of the two-slit experiment \cite{DEWD, QPPLOTS, K89} it is seen that trajectories  never cross so that a particles path can be theoretically determined with certainty even when their is interference, and irrespective of whether the experiment is of an intermediate type or not.}. The visibility, as in classical physics, is defined with respect to a 100\% wave concept, and measures the coherence of a particular wave profile, i.e., a particular profile of the $R$ and $S$-fields.

\section{The Particle-Wave Duality Relation\label{SECPWDR}}
An indication of the importance of the WZ analysis is indicated by the interest it has attracted from numerous authors, who, based on the WZ analysis developed a particle-wave duality relation. The first such relation was introduced by Greenberger and Yasin (GY) \cite{GY88}, with further developments in references \cite{JSV95, ENG96, SM89, ZWM91,  M91,  LHGZ2012}. Among the latter are detailed derivations, the most notable of which are by Jaeger, Shimony and Vaidman (JSV)\cite{JSV95}, and then by Englert \cite{ENG96}. The JSV derivation is notable because of the detailed discussion on the interpretation given to the duality relations. The Englert derivation is notable because of its mathematical rigor.

Although the various authors beginning with Greenberger and Yasin arrive at a similar expression of the particle-wave duality relations, 
\begin{equation}
D^2+V^2\leq1,\label{PWDR}
\end{equation} 
and to a similar use  of $D$ as a measure of the  particle aspect, and of $V$ as a measure of the wave aspect, the precise interpretation  given to $D$  varies considerably. Later authors define $D$ as the distinguishability, while $V$ is  always  defined as the visibility. The particle-wave duality relation, given that it originates from WZPC, suffers from the same criticisms as given above for WZPC. Like WZPC, the  duality relation can be given meaning with reference to an ontology such as the  causal interpretation, but in this case  its conceptual significance is lost. Since in this interpretation  fundamental entities are modeled as particles guided by fields, it becomes meaningful to ask for the probability that a particle passes through one slit or the other in a two-slit arrangement. It also becomes  meaningful  to ask for the visibility of the interference fringes, since the profile of the fields (the $R$ and $S$-fields) guiding  the particles determine the visibility. Making one slit half the size of the other will decrease the visibility while increasing the probability of a particle passing through the larger slit. 

But note, in the causal interpretation, the duality relation is {\bf not} interpreted as ``how wave-like'' or ``how particle-like'' a quantum system is. Particles are 100\% particles and fields (waves) are a 100\% fields (waves). Instead, definite probabilities can be attached to each path a particle might take (from which the path parameter $P$ can be defined). The visibility $V$ is determined by direct measurement of the maximum and the minima of the interference pattern, but there is no question of the level of visibility placing a limit on the wave concept; the full wave concept is needed to explain even the smallest level of fringe visibility.

But, a more correct model of photons is given by the causal interpretation of the electromagnetic field (CIEM) \cite{K06, K94, PNK2005}. In this interpretation there are no photon particles; photons are fields (waves). With reference to this ontology, the duality relations must be interpreted entirely differently, perhaps as a measure of the coherence of a particular wave profile, without reference to particles.  

Without reference to an ontology the interpretation of the duality relations requires care to avoid ambiguity or even contradiction. If the duality relations are interpreted along the lines of WZPC, i.e., as a measure of partial wave behaviour or partial particle behaviour, then again we are faced with a contradiction of definitions.  Further, the use  of the visibility as a measure of the wave aspect of a physical system seems questionable, since even the tiniest level of interference requires a full wave model for its explanation. If instead, the duality relations are interpreted in terms of particle and wave information only, contradiction might be avoided, but such an interpretation leaves the question of what the information is referring to unanswered. This is hardly satisfactory. We surely expect some kind of explanation of how an experimental result comes about.  Since both of the latter alternative interpretations apply the particle and wave concepts to the same experiment, they are not consistent with BPC. The best interpretation that does not refer to an ontology, in my opinion, is due to JSV \cite{JSV95}. They interpreted the duality relation in terms of predictability with the path parameter and the visibility parameter referring to two different mutually exclusive experimental arrangements\footnote{Drezet also makes this point in his articles given in ref. \cite{DREZ}. In the second of these articles he discusses the duality relation in some detail in the context of arguing against  Afshar's claimed experimental refutation of BPC.\label{DREZFT}}. Thus, the duality relation predicts that if in a given experimental arrangement (e.g. the two-slit arrangement) the path is measured with result $P$ for the path parameter, then a separate mutually exclusive measurement of visibility $V$ would yield a result  consistent with the duality relation (\ref{PWDR}). Thus, by interpreting the particle-wave duality relation as referring to two mutually exclusive experimental arrangements, consistency with BPC is achieved.

We see that the duality relation can be interpreted in different ways depending on the point of view of the interpreter. This fact, especially because of the existence of  ontological interpretations, in my view, diminishes the fundamental significance normally attributed to it.

\section{What is a measurement?\label{WISAM}}

Although it is not our purpose here to discuss the measurement theory of quantum mechanics, we will include a brief overview, including a brief outline of Bohm's proposed solution of the measurement problem \cite{B52} pp. 180 - 193, which is my preferred solution. The particular questions we want to address, since their answer is central to our defense of BPC are, ``What constitutes a measurement?'', and ``What constitutes an inference, based on an assumption or assumptions, drawn from the experimental arrangement?''

A detailed mathematical treatment of measurement was given by von Neumann in his book \cite{N55}. Von Neumann's analysis shows that a  measurement results in an entangled state consisting of a sum of terms, with each term consisting of an apparatus state correlated to  an eigenstate of the observable being measured. However, what is observed after the measurement is not the entangled state, but the reduction of the entangled state to a single term representing a single apparatus state (pointer position) and its correlated  eigenstate. This reduction is not described by the Schr\"{o}dinger equation, i.e., it  is not described by the quantum theory. This is the measurement problem of the quantum theory. Some authors characterise the problem by considering that what is observed is not the entangled state resulting from solving the Schr\"{o}dinger equation, but a mixed state, e.g. \cite{London}.

To date, consensus as to the solution of the measurement problem has still not been reached. A number of solutions have been proposed, and remain in contention. Bohm, for example, gives a detailed mathematical treatment of measurement  and tries to answer the measurement problem by suggesting that the apparatus introduces random phase factors which destroy interference between the  different terms of the entangled state. But, this approach does not explain why a particular term is singled out \cite{B51}, ch. 22. Later, Bohm, based on the causal interpretation, gave another solution of the measurement problem \cite{B52} pp. 180-193, which, as mentioned earler, I personally favour and will outline below. For other approaches  to the question of measurement in the quantum theory we direct the reader to the well known book by Wheeler and Zureck,  \cite{W83}. A particularly nice article  from this book is by London and Bauer \cite{London}. Aside from Bohm's proposed solution, the main other proposals are: Heisenberg's \cite{H52a, H52b} and Von Nuemann's  \cite{N55} introduction of an arbitrary classical-cut separating the quantum and classical regimes; that the collapse of the wave function occurs in the consciousness of the observer has been suggested by a number of authors, e.g. London and Bauer \cite{London}, von Nuemann \cite{JAM74} pp. 474 Sect. 11.2 and notably by Wigner \cite{WIG76}; Everett's many-worlds (relative-state) interpretation \cite{EVER57}; and the decoherence approach \cite{ZEH70, SCH2005}, extensively developed by Zureck \cite{ZUR91,ZUR2003}. All of these approaches are thoroughly discussed in the references given, as well as elsewhere, so we will restrict ourselves to a few comments.

That each term of the entangled, often infinite, series resulting from  a measurement corresponds to a separate universe is fanciful indeed. Many natural atomic and sub-atomic interactions are essentially measurements, which may be called natural measurements. So, in Everett's suggested solution  we are supposed to believe that of the enormous multitude  of natural and man-made measurements on atomic and subatomic systems,  each results in the creation of infinite numbers of new universes, and that in each of these new universes, the  enormous multitude  of natural and man-made measurements of atomic and subatomic systems, themselves each produce further infinite numbers of universes. This proposed solution certainly lacks conceptual economy, and for many, like myself, the idea of the continuous creation of infinite multitudes of universes, especially in the absence of the slightest evidence, is totally unacceptable. For example, this proposal  begs the mind-boggling question: ``In  what kind of super space-time do all  these universes exist?''. A more substantial  criticism  is the fact that the terms of the entangled state resulting from measurement are not instantaneously orthogonal. During the interaction time of the system with the apparatus there is a substantial overlap of terms leading to a very complicated, and not necessary small, upheaval. Typically, the apparatus in the  mathematical treatment of measurement is initially  represented by a wave packet (e.g. a Gaussian). For this initial state of the apparatus the solution of the Schr\"{o}dinger equation with the appropriate interaction term $H_I$ is a series of apparatus packets, each correlated to an eigenstate of the observable being measured. It is only when these apparatus packets (correlated to an eigenstate of the observable being measured) separate, i.e., become approximately orthogonal, that a measurement can be said to take place. During the interaction interval, therefore, the wave packets overlap, i.e., interfere with each other, with effects that may be very large. These effects can be thought of as ripples in the universe. During the interaction interval and before the packets separate, these, not necessarily small, ripples should be observable in the universe an observer happens to be in every time a measurement (natural or man-made)  takes place. No such ripples have  ever been observed.

The notion that the collapse of the wave function occurs in the consciousness of an observer is also subject to series objection. The idea arises from von Neumann's analysis of measurement, mentioned above, in which the apparatus is treated quantum mechanically. When does the classical observation of a single term of the entangled series come about, i.e., when does the collapse of the wavefunction occur? One can arrange a second apparatus to measure the state of the first. But again, what is the state of the second apparatus? One can then employ a third apparatus to observe the state of the second apparatus and so on. This leads to von Neumann's infinite regression. The process must stop somewhere, and this gives rise to the idea of the classical-cut (which we will discuss in more detail later) where suddenly quantum laws cease to function and classical laws take over. The arbitrariness of the position of the cut is a serious objection to the notion of a classical cut. That the collapse occurs in the consciousness of the observer is a proposed solution to the problem of where to place the classical cut.  The implication of this idea is that a macroscopic apparatus remains in a superposed state until such time that it is observed by an observer. Then, what gives the observer the  special role of singling out only one term of the entangled series. If the macroscopic apparatus is governed by the laws of quantum mechanics, why not the human observer? Why does the observer not  enter into the entangled state, with each state of the observer correlated via the apparatus to an eigenstate of the operator being measured? One can argue that the observer knows that he is in a single state. But, to another observer the entangled state of system+apparatus+observer exists until such time as he makes his observation.  In other words, the issue is not only the conventional one raised in many texts, namely, how far into the consciousness should the cut be placed. Rather, it extends outside of the single observer and becomes an issue of an infinite chain of observers. Clearly, this solution is untenable.

In the case of Wigner, the idea of collapse occurring in the consciousness is tied to the notion that the wave function does not have objective reality but rather represents a probability, i.e., quantum mechanics is a probability calculus \cite{WIG76} (recall also my earlier suggestion in Sect. \ref{BPC28}, footnote \ref{FF2}, that Bohr may also have held this view). The notion of collapse occurring in the consciousness of an observer then becomes an issue of acquisition of more knowledge by the observer, leading to a representation of the new knowledge by only a single term of the entangled series. In this form the idea of collapse produced in the consciousness of the observer becomes more palatable, but if quantum mechanics is to be viewed as a probability calculus, why introduce collapse or involve consciousness at all? If the wave function is not attributed objective physical reality and if it is viewed as only providing probabilities, then the issue of collapse and therefore a measurement problem does not arise (it can even be argued that in this view nonlocality doesn't arise). However, this view raises the new question, ``What brings about the probabilities?'', i.e., "What underlies the probabilities?" Classical probability is underpinned by an ontology governed by causal law. Surely, quantum probabilities do not come about by magic, but rather there are underlying processes, whether we understand them or not (at present),  that give rise to quantum probabilities.  I would like to pursue the idea of quantum mechanics as a probability calculus without objective reality a little further as I think it can lead to a world view that is consistent with BPC that removes the need for the rather undesirable (in my view physically inconsistent) classical cut.

I present this view because I believe it to be self-consistent even though it is not my preferred view. As mentioned earlier my preferred view is the causal Interpretation. My starting point is the following observation. The entire quantum theory, from its very beginnings, is built from results obtained from macroscopic apparatus. That the results of many experiments were not consistent with classical theory does not change the fact that all of these results were obtained from macroscopic apparatuses. From this observation we can conclude that the relation of the classical and quantum worlds is inbuilt in the quantum theory.  The implication of this is that the transition from the quantum to the classical world is a smooth transition, whether  or not we understand the mechanism of the transition (Actually,  I think that Bohm's answer  to the measurement problem based on the causal interpretation and decoherence  provides a mechanism for this transition and I will discuss this later, but this point is not needed for our present argument). With this in mind, I present the following world view consistent with BPC and which avoids the measurement problem: Quantum theory is a probability calculus, i.e., the wave function does not have an objective reality. But, there  is an objectively existing underlying physical reality which gives rise to the quantum probabilities observed by macroscopic apparatus, but which, according to BPC, is completely and permanently beyond our conceptual gaze. Moreover, the underlying physical processes work in such a way to produce the classical world. In this view, there is no measurement problem, no need for a classical cut, and no need to invoke the consciousness of the observer in the description of measurement.

The first suggestion of decoherence  is typically attributed Zeh \cite{ZEH70}, who explicitly introduced the role of the environment in the solution of the measurement problem. But, if one reads Bohm's proposed solution of the measurement problem based on the causal interpretation, decoherence and the environment also play an essential role.  In Bohm's solution, the decoherence arises from the many thermodynamic degrees of freedom of a macroscopic apparatus, or more accurately, the inevitable coupling of the apparatus variable/s with the many thermodynamic degrees of freedom of the environment (because of processes such as friction, Brownian motion, etc.). I wonder though, why the environment has to be brought in separately, when the important requirement for decoherence is the involvement of many degrees of freedom, and any macroscopic apparatus at some stage of its operation involves many degrees of freedom.

Let me now proceed to outline Bohm's solution to the measurement problem (for details we direct the reader to  \cite{B52} pp. 180-193), with the aim of emphasising two important features: First, that  decoherence plays an essential role in the solution of the measurement problem. Second, that the transition from the quantum to the classical world is gradual. We note that Bohm's  treatment of the measurement problem can be extended to boson fields by considering a combined space consisting of ordinary space and the space of normal mode coordinates. The essential features of the causal interpretation  that we will need  are that quantum objects (electrons, protons etc.) are particles guided by two fields, the $R$ and $S$-fields defined by   $\psi(x,t)=R(x,t)e^{(iS(x,t)/\hbar)}$. The $R$ and $S$-fields are not independent, but codetermine one another, meaning that the motion of a particle can be determined from either field. Now, consider a measurement of an observable $Q$, with eigenvalue equation $Q\phi_q(x)=q\phi_q(x)$, of an electron, say,  described by the initial  wave function $\psi(x,t=0)$, using an apparatus described by the initial wave function $g(y,t=0)$, where $y$ can represent a single or many variables. Since initially, the apparatus and electron are independent, the initial combined wavefunction is the product
\[ 
\Phi(x,y,t=0)=\psi(x,t=0)g(y,t=0).
\]
Generally, $g(y,t=0)$ will have the form of a wave packet (e.g. a Gaussian wave packet) since the apparatus always has a definite position (at least within some well defined interval $\Delta y$). For example, $g(y,t=0)$ may represent the initial position of an instrument pointer \footnote{We will not here take up the question of whether or not all measurements can ultimately be reduced to position measurement, but assume that this is the case.}.

Bohm considers an impulsive measurement, i.e., a very strong interaction lasting for a very short time. The action for only a very short time allows the changes in the apparatus and electron wave functions that would take place during the interaction time to be ignored. This allows the mathematical simplification of neglecting the individual Hamiltonians of the electron and apparatus, leaving  only the  interaction Hamiltonian $H_I$ in the  Schr\"{o}dinger equation, i.e.,
\begin{equation}
i\hbar\frac{\partial \Phi(x,y,t)}{\partial t}=H_I\Phi(x,y,t).\label{SCEE}
\end{equation}
In order for the system and apparatus to be coupled, $H_I$ must depend on operators that commute with  $Q$ \footnote{According to Bohm this requirement is needed so that $H_I$ does not produce any uncontrollable changes in the observable $Q$, but only in observables that do not commute with $Q$. However, according to quantum theory, an arbitrary initial  wave function cannot have a well defined eigenvalue of $Q$ prior to measurement, unless it is an eigenstate of the observable being measured. It is the measurement process that changes the wave function into an eigenfunction of $Q$, with a corresponding definite eigenvalue $q$. In general, therefore, both the wave function and the value of $Q$ is changed by a measurement. If this were not the case one could equally well envisage a mutually exclusive 
 impulsive measurement that does  not change the value of an observable $P$ that  does not commute with $Q$. This would imply that both $Q$ and $P$ have well defined values prior to measurement, whereas the wavefunction does not describe these values. This would  constitute a variant of the EPR incompleteness argument. We conclude that $Q$ and $P$ cannot have definite values prior to measurement, unless the wave function is an eigenstate of one of them, in which case, only that observable will have a definite value.},  and on operators that depend on $y$. A simple choice of $H_I$, for example, is $H_I=-aQp_y$, where $a$ is a constant and $p_y$ is the momentum operator conjugate to $y$. 

The solution of eq. (\ref{SCEE}) has the form 
\begin{equation}
\Phi(x,y,t)=\sum_{q}\phi_q(x)g_q(y,t),\label{SOLNN}
\end{equation}
which is an entangled series in which each term represents a correlation between the apparatus state (pointer position) and an eigenstate of the observable $Q$. Since initially $g(y,t=0)$ has the form of a wave packet, so also will $g_q(y,t)$. During interaction the wave packets overlap, so the system and apparatus undergo very complicated  motions. As the packets separate, the combined system begins to quieten down, until after a sufficiently large separation only one wave packet governs the behavior of the apparatus, while the correlated eigenstate governs the behaviour of the quantum system. The packets have infinite tales and so never truly separate, but for sufficient separation affects produced by the overlap of the tails are so negligible  that they can be neglected for all practical purposes. The probability of the packets overlapping again after they have separated when very many degrees of freedom are involved may be compared to the probability of all the molecules in a room gathering  on one side, leaving a vacuum on the other side. Though possible, it is so overwhelmingly unlikely, that such an ordering of molecules in a room has never been observed. Similarly, the reversal of a macroscopic measurement has never been observed. 

Once the packets have separated only the $R$ and $S$-fields of the wave packet (and eigenstate) located at the position of the apparatus (and system) influence the apparatus (and system), i.e., only one term of the series is ``active''. The tails of the other wave packets have a negligible affect at the location of the active packet. Similarly, since all other eigenstates are orthogonal to the active eigenstate, they also will not affect the system or the apparatus. Since the amplitude of the wave packets in the space between their peaks  is negligible, the probability $|\Phi(x,y,t)|^2$ between the packets is also negligible, so that once the apparatus enters a particular wave packet  and the system enters the correlated eigenstate, both remain governed by these thereafter. We see that it is the  ontology of the causal interpretation that resolves the measurement problem without collapse, since, in addition to the apparatus having a definite position,  quantum objects,  being  particles, also have a definite position,  and it is the  position of the apparatus and quantum system that selects the ``active'' packet. The active term can be renormalised without affecting any physical quantity, so that for all practical purposes the state of the apparatus and system will be given by
\[
\Phi(x,y,t)=\phi_q(x)g_q(y,t).
\]
This explains why only one term of the entangled series solution of the Schr\"{o}dinger equation is observed in a measurement without the need to invoke the  collapse mechanism, a mechanism not described by the quantum theory.

The apparatus variable $y$ may represent one thermodynamic degree of freedom, i.e. a single variable, (as is the case of an atomic coordinate in a Stern-Gerlach experiment), a few degrees of freedom or a myriad of thermodynamic degrees of freedom. In the first two cases the wave packets can be made to re-overlap either preventing a definite measurement or reversing the measurement, i.e., restoring the system to its original state (we will give an example of the latter below).  Where many thermodynamic degrees of freedom are involved, the possibility to overlap again once they have separated becomes overwhelmingly unlikely, since the packets will fail to overlap if they are widely spaced for even a single coordinate. The possibility for the  wave packets to re-overlap once they have separated can, therefore, be neglected for all practical purposes. In this case, an irreversible measurement has been achieved. Ultimately,  any measurement, the result of which can be observed by an experimenter, falls into the latter category since observation requires amplification to the macroscopic level,  which  always involves a myriad of thermodynamic degrees of freedom. Bohm emphasized this,  pointing out  that for any real measurement the apparatus variable eventually must become coupled to many degrees of freedom (e.g. through friction in the pointer pivot etc.). Thus, Bohm's proposed solution of the measurement problem involves the environment and precedes Zeh's suggestion.  We may note that including the environment separately, i.e.,  writing the solution of eq. (\ref{SCEE}) as
\[
\Phi(x,y,t)=\sum_{q}\phi_q(x)g_q(y,t)e_q(z,t),
\]
seems to be essentially equivalent to the solution (\ref{SOLNN})  of eq.(\ref{SCEE}), when the apparatus wave packet is a function of many degrees of freedom.

Finally, we would like to emphasize, as promised earlier, that Bohm's proposed solution indicates how the transition from the quantum to the classical world is a gradual one. As long as only a few degrees of freedom are involved quantum affects are prominent, but as more and more thermodynamic degrees of freedom come into play quantum affects are increasingly suppressed until at the macroscopic level quantum affects are negligible and classical laws serve as extremely good approximations.

In passing, though not relevant to the present discussion, we mention that Aharanov et al  introduced new measurement protocols, namely protected measurements \cite{AV93, AAV93}, and weak measurements, first introduced in 1988 \cite{AAV88}, and developed further in 1990 \cite{AV90}. Interest in weak measurements is gaining momentum \cite{AS2013}, with weak measurement experiments performed in connection with  Hardy's paradox \cite{LS2009,YYKI2009}. The defining property of a protective measurement is that the system returns to its initial state.  A weak measurement is made by pre-selecting a state (i.e. preparing a system in a given initial state $| i \rangle$), applying a weak interaction (e.g. a weak magnetic field), then post-selecting a  state $| f \rangle$. Measurements are repeated for a large number of systems in the same pre and post-selected states.  Mathematically a weak measurement is defined by
\[
\langle A \rangle=\frac{\langle f |A| i \rangle}{\langle f| i \rangle}.
\] 
Weak measurements allow a ``picture'' to be painted of quantum systems between measurements. However, it is clear that the pictures are averages over a large number of systems in the same pre and post-selected states, so that it is by no means clear that the pictures so obtained  correspond to actual quantum reality, i.e. the pictures so obtained may be entirely fictitious. Here it is worthwhile to recall Bohr's caution, ``...the position of an individual at two given moments can be measured with any desired degree of accuracy; but if, from such measurements, we would calculate the velocity of the individual in the ordinary way, it must be clearly realized that we are dealing with an abstraction, from which no unambiguous information concerning the previous or future behaviour of the individual can be obtained.''   \cite{Bohr34} pp. 66 and \cite{BR28} (b) pp. 583. Yet, experimentalist suggest that the concept of weak measurement has tremendous technological promise  \cite{ZUR91}.

We come now to answer the questions posed at the beginning of this section, namely, ``What constitutes a measurement?'', and ``What constitutes an inference, based on an assumption or assumptions, drawn  from a particular experimental arrangement?'' This distinction is important for our defense of BPC. The answer depends on a particular ontology or a particular epistemology. From the perspective of Bohm's ontology what constitutes a measurement is answered above, but a little more elaboration and emphasis is needed. In Bohm's ontology a system prepared in a particular state may be considered a measurement even before it interacts with a macroscopic apparatus and before any observer is involved, since the state of the system is known with certainty as can be confirmed by a subsequent interaction with a macroscopic apparatus\footnote{This differs from the EPR criteria for elements of  reality, since in the EPR experiment the system is not an eigenstate of either of the non-commuting observables whose values are predicted with certainty.}. 

For Bohr's epistemology, i.e.,  BPC, on the other hand, a measurement is required to be amplified irreversibly and recorded on a macroscopic apparatus. Wheeler goes further, as expressed in his famous quote, ``No phenomenon is a phenomenon until it is an observed phenomenon. '' \cite{WHR78} pp. 14. Two absolutely essential, though connected, features define a Bohr measurement: (1) Amplification to the macroscopic level, and (2) Irreversibility. The irreversibility is invariably a consequence of amplification. It is important to note that a null result also constitutes a measurement as long as the possibility of making a macroscopic, amplified, irreversible record is there. This latter type of measurement is what is performed in the Afshar experiment in determining the existence of an interference pattern as we shall explain in Sect. \ref{AFECRT}. Such a null measurement may be considered a type of non-demolition measurement. Bohr never gave a precise criteria for what constitutes a macroscopic apparatus, but rather he considered an apparatus to be macroscopic when it is sufficiently heavy. We may take this working definition to be essentially equivalent to an apparatus being considered macroscopic when it contains a sufficient number of degrees of freedom. 

Since our intention is to defend BPC we will define an ``inference'' as opposed to a measurement in the context of BPC: An inference is a conclusion drawn from an experimental configuration which neither involves an actual macroscopic, amplified, irreversible record, nor is there the possibility to make such a record (to account for null measurements). Thus, an ``inference'' is neither a measurement, nor can it be considered a null (nondemolition) measurement. The distinction between a measurement  and an inference is therefore characterised by the absence of either a macroscopic, amplified, irreversible record, or the possibility to make such a record . Instead, an inference drawn from an experimental configuration depends on one or more assumptions.

We may illustrate the above with an example of a spin-1 measurement using the Stern-Gerlach setup described by Feynman  \cite{F64}, vol. III, ch. 5, to which the reader is referred for details and diagrams. 

A beam of atoms of spin $l=1$ and arbitrary orientation is passed through an inhomogeneous magnetic field of width $w$, and increasing gradient $\nabla \vec{B}$ in the positive $z$-direction. For sufficient $w$ and strength, the beam will split into three beams of spin $l=1 $ and orientation $m_z=-1$, 0 and 1. In this case the apparatus variable is the coordinate of the atom. By blocking the $0$ and $-1$ beams, the transmitted beam is known to be in an $l=1$, $m_z=1$ spin state with certainty, as can be confirmed by recording the positions of the atoms in the beam on a detecting plate (which conforms to a measurement as defined above). Before the detecting plate, the spin state of the atoms in the beam is known with certainty,  but no macroscopic, irreversible amplification has taken place nor is an observer involved. In Bohm's ontology this constitutes a measurement, but in BPC this is not a measurement until the atoms position is recorded on a detecting plate. i.e., until a macroscopic, amplified, irreversible record is made.  Since only a single apparatus variable is involved before a detecting plate, the measurement can be reversed, i.e., the atoms can be returned to their original spin state, by transmitting all three beams through a second magnetic field of width $2w$ with opposite orientation to the first, then through a third magnetic field identical to the first.\\

\section{Optical Experimental Tests of Complementarity}

Having - hopefully - clarified aspects of complementarity, the particle-wave duality relations and measurement, I come  to my main focus which is to argue that quantum  optical experiments which claim to refute complementarity fail in their aim. As mentioned, I will consider the GHA experiment  \cite{GHA91}, the BGGP experiment  \cite{BGGP04}, and especially the Afshar experiment  \cite{AFSHAR04}, which has received a great deal of attention. Some of the experiments test the duality relation and not BPC directly. As I mentioned earlier, despite this, all experiments ultimately test BPC. Since the same reasoning saves both BPC and the duality relations,  the experiments need not be considered as separately testing BPC and the duality relation. 

A refutation of complementarity requires two actual measurements (as defined in Sect. \ref{WISAM}) in the same experiment, with one measurement consistent with the wave concept, while the other measurement is consistent with the particle concept. For example, one of the measurements should be an interference pattern recorded indelibly  on a detecting plate ``click'' by ``click'', while the other measurement should be a definite indelible  record in a path detector. Yet, to repeat what I have said in the introduction, in all of these claimed experimental refutations, one complementary concept is actually measured (i.e. defined by the experimental arrangement as required by BPC), while the other complementary concept is artificially introduced through an unjustified assumption or assumptions. In such experiments, the experimenters include some intermediate process, which in classical theory must  be described by the  classical concept opposite to the complementary classical concept consistent with final actual measurement. But, since this concept is not actually measured, it is not even defined in the way demanded by BPC. Thus, BPC (and even the duality relations) are not even nearly challenged by these experiments. 

I strongly suspect that an experiment which enables the simultaneous measurement of complementary concepts  (whether particle-wave concepts or canonically conjugate variables), will also contradict the quantum formalism itself.
 
\section{The Experiment of Ghose, Home and Agarwal}

Here I consider the  1991 experiment proposed by Ghose, Home and Agarwal \cite{GHA91}  and later performed by Mizobuchi and Ohtak\'{e} \cite{MO92}. It is based on the experiments performed by Bose in 1887  in which Bose observed the tunneling of microwaves when two asphalt prisms were placed sufficiently close to each other,  \cite{B27, S64} and \cite{F64} vol. II, pp. 33 -12. The novelty of the GHA proposed experiment is the use of  single photon states. 

Genuine single photon states, or one photon Fock or number states, are not as easy to produce as might be thought.  Early experiments used very low intensity chaotic light where on average only one photon at a time is present in the apparatus. But, such low intensity light is not a Fock state and possesses very different properties. As I mentioned earlier, Grangier, Roger and Aspect were, perhaps, the first to use genuine single photon states. Pairs of entangled photons produced either by atomic cascades or by parametric down conversion are used. One of the pair is used to trigger  a gate which remains open for a time of suitable length to allow its photon partner to pass through a gate with a very high probability. The photon partner is then in a single photon state (at least to a good approximation). 

\begin{figure}[h]
\unitlength=1in
\hspace*{1in}\includegraphics[width=2.8in,height=2in]{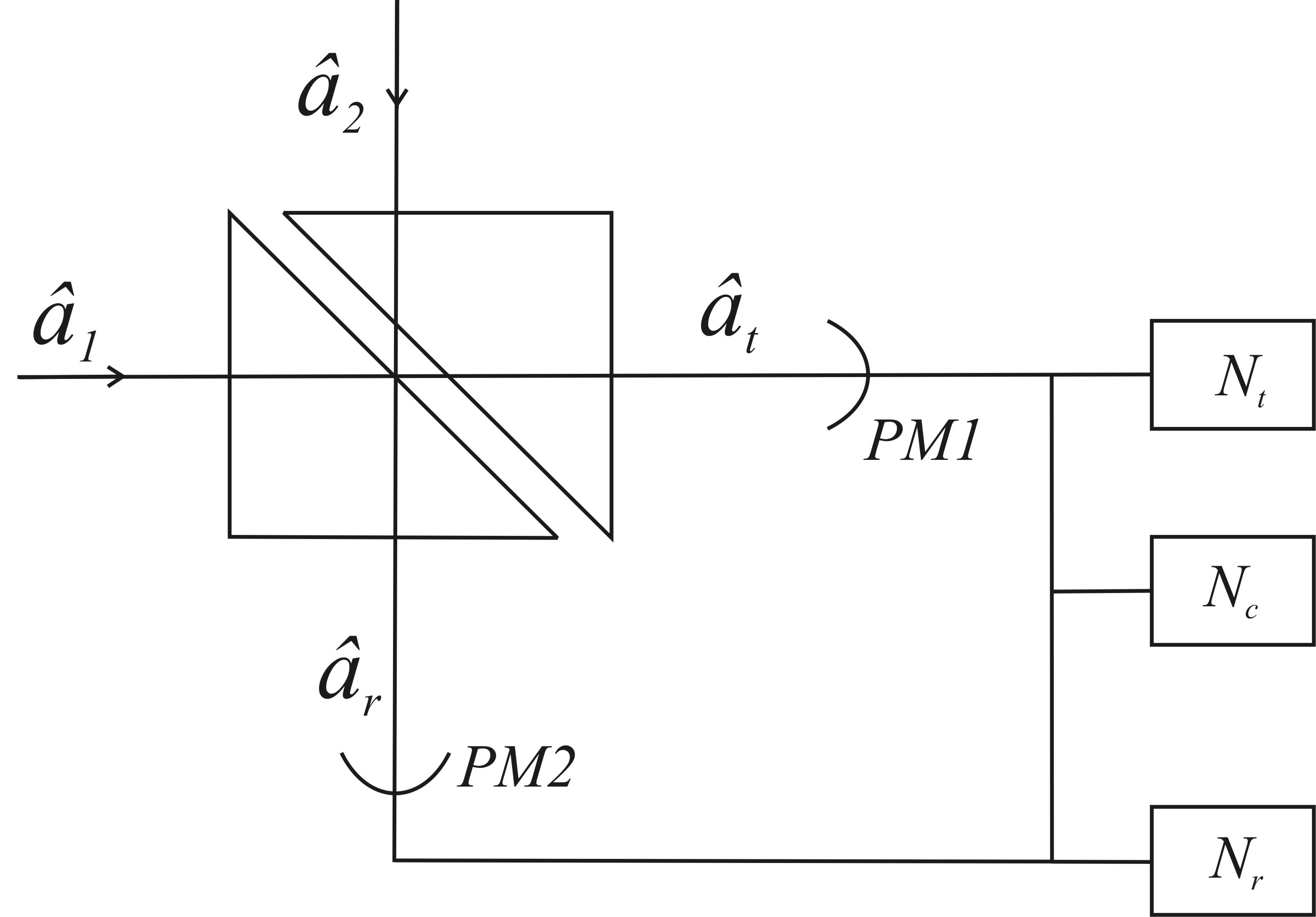} 
\caption{ GHA's beam-splitter configuration using a single photon source. When the gap is of the  order of a wavelength, tunneling can occur.}
\label{fig1}
\end{figure}

Consider the configuration shown in figure \ref{fig1}. The single photon source is arranged so that only one photon is present in the apparatus at any one time. Since the $45^\circ$ angle of incidence is greater than the critical angle, total internal reflection occurs  when only one prism is in place. In this case, we expect  only photomultiplier $PM1$ to register a series of single photon counts. When the gap between the two prisms is reduced to about a wavelength, tunneling can occur, so that a photon may be either reflected or transmitted.  

This two-prism configuration can obviously be treated mathematically in the same way as in the case of a single prism beam splitter \cite{ZEIL81, CST89, LOUD2000}.  Let $R$  be the reflection coefficient and $T$ the transmission coefficient. There values will depend on the gap between the two prisms. Phase changes are included by allowing $R$ and $T$ to be complex numbers. It is  assumed that the two prism configuration corresponds to a lossless symmetrical beam splitter. 

In figure \ref{fig1}, $\hat{a}_1$ and $\hat{a}_2$ are the input  annihilation operators, and $\hat{a}_r$ and $\hat{a}_t$ are the output annihilation operators. These are related by
\begin{equation}
\left(\begin{array}{c}
\hat{a}_r\\
\hat{a}_t
\end{array}\right)
=
\left(\begin{array}{cc}
R&T\\
T&R
\end{array}\right)
\left(\begin{array}{c}
\hat{a}_1\\
\hat{a}_2
\end{array}\right),
\end{equation}
or
\begin{eqnarray}
\hat{a}_r&=&R\hat{a}_1+T\hat{a}_2,\label{ar}\\
\hat{a}_t&=&T\hat{a}_1+R\hat{a}_2.\label{at}
\end{eqnarray}
The creation and annihilation operators satisfy the usual commutation relations
\begin{equation}
[\hat{a}_i,\hat{a}^{\dag}_j]=\delta_{ij},
\end{equation}
where $i,j=1,2,3,4$. These commutation relations led to the following conditions on $R$ and $T$
\begin{eqnarray}
|R|^2+|T|^2&=&1,\label{R}\\
RT^*+TR^*&=&0.\label{T}
\end{eqnarray}
In the  experiment, a single photon enters the beam-splitter from input 1, while input 2 is in the vacuum state. Hence, it is necessary to find $\hat{a}_1^{\dag}$, which can be done using eq.'s (\ref{ar}) to (\ref{T}):
\begin{equation}
\hat{a}_1=R^*\hat{a}_r+T^*\hat{a}_t,\label{a1}
\end{equation}
The complex conjugate of eq. (\ref{a1}) gives 
\begin{equation}
\hat{a}^{\dag}_1=R\hat{a}^{\dag}_r+T\hat{a}^{\dag}_t.
\end{equation}
Let $| 0 \rangle=| 0 \rangle_1| 0 \rangle_2=| 0 \rangle_r| 0 \rangle_t$ be the vacuum state for all the beam-splitter inputs and outputs. The state before and after the beam splitter may now be written as: 
\begin{eqnarray}
\hat{a}^{\dag}_1| 0\rangle&=& | 1\rangle_1| 0\rangle_2\;\;\;\;\;\;\;\;\;\;\;\;\;\;\;\;\;\;\;\;\;\;\;\;\;\;\;\;\;\;\;\;\;\;\;\nonumber\\
&=&R\hat{a}^{\dag}_r|0\rangle_r|0\rangle_t+T\hat{a}^{\dag}_t |0\rangle_r|0\rangle_t\\
   &=& R|1\rangle_r|0\rangle_t + T|0\rangle_r|1\rangle_t.
\end{eqnarray}
The state after the beam-splitter is seen to be an entangled state. The amplitude $A_r$ for reflection is   
\begin{eqnarray}
A_r&=& _r\langle 1|_t \langle 0| 1\rangle_1| 0\rangle_2\nonumber\\
&=&R _r\langle 1|1\rangle_r  \mbox{}\;_t \langle 0|0\rangle_t  +  T_r\langle 1|0\rangle_r\mbox{}\;  _t \langle 0|1\rangle_t\nonumber\\
& =&R,
\end{eqnarray}
 and the amplitude $A_t$ for transmission is
\begin{eqnarray}
A_t&=&  _r\langle 0|_t \langle 1| 1\rangle_1| 0\rangle_2\nonumber\\
&=&R_r\langle 0|1\rangle_r  \mbox{}\;_t \langle 1|0\rangle_t  +  T_r\langle 0|0\rangle_r\mbox{}\;  _t \langle 1|1\rangle_t \nonumber\\
&=&T.
\end{eqnarray}
The corresponding probabilities  are $|A_r|^2=|R|^2$ and $|A_t|^2=|T|^2$. The amplitude  $A_c$ for a coincidence count is 
\begin{eqnarray}
A_c&=&  _r\langle 1|_t \langle 1| 1\rangle_1| 0\rangle_2\nonumber\\
&=&R_r\langle 1|1\rangle_r  \mbox{}\;_t \langle 1|0\rangle_t  +  T_r\langle 1|0\rangle_r\mbox{}\;  _t \langle 1|1\rangle_t\nonumber\\& =&0,
\end{eqnarray} 
so that the  probability $|A_c|^2$ for a coincidence count is also zero.

GHA took perfect anticoincidence to be consistent with particle behaviour. That is to say, the final experimental results are consistent with particle behaviour. For a photon detection in PM1 tunneling must have occurred.  Classically tunneling at the gap can only be explained by wave theory. Therefore, GHA argued that the  classical requirement that tunneling must be explained by the wave concept amounted to the observation of wave behaviour. GHA therefore concluded that wave and particle behaviour are observed in the same experiment in direct contradiction to BPC. In this way, GHA claimed that their experiment constitutes a refutation of  BPC. The theoretically predicted results were confirmed by the experiment of Mizobuchi and Ohtak\'{e}. This was expected as different results to those predicted would contradict quantum theory.  

The refutation however fails. The key point is that the wave behaviour is never actually measured, but rather, it is only inferred based on the circumstance that classically tunneling can only be explained in terms of the wave concept. However, Bohr insisted that a description of mechanisms underlying experiment is impossible  and that classical concepts  are in any case abstractions to aid thought that cannot be attributed physical reality. The only requirement of BPC is that the classical concept be consistent with the experiment as a whole, i.e.,  the experimental arrangement and the final experimental result. Indeed, it is the experimental arrangement and measurement that defines the concept.  Since the wave concept is inferred but not measured it is completely illegitimate under BPC to apply the wave concept to the GRA experiment. Thus, according to BPC, the GHA experiment is consistent with (defines) one and only one complementary concept, namely,  the particle concept. I conclude that the GHA experiment is perfectly consistent with BPC.

To complete our refutation we need to justify why particle behaviour is consistent with a measurement, while wave behaviour is not. First,  the ``clicks'' in the photomultiplier tubes revealing the photon's path fulfill the definition of measurement given in Sect. \ref{WISAM}, since the photon initiates the release of a cascade of electrons producing a current which is recorded on a pointer (or digital readout), a process which constitutes a macroscopic, amplified, irreversible record. The detections in photomultiplier $PM1$ certainly confirm transmission, but do not reveal by what mechanism the photons jumped the gap, whether they were transmitted as waves or whether they were particles that jumped the gap. In the process of transmission there is no macroscopic, amplified,  irreversible record of any sort and certainly no such record that is consistent with either the wave or particle concept, nor is there any possibility for such a record to be made. We conclude that transmission through the gap niether fulfils the conditions for a measurement,  nor for a null measurement as they are defined in Sect. \ref{WISAM}, and therefore transmission measures neither wave nor particle behaviour. It follows that the claim by GHA that transmission through the gap confirms wave behaviour is an inference based on assumption and not a measurement of wave behaviour.

That BPC is consistent with GHA  does not mean that BPC provides a satisfying description of the experiment. Many physicists, including the present author, want to picture the physical mechanisms leading to the final experimental result. However, borrowing classical explanations in an ad hoc way is itself not very satisfactory. It is for these reasons that the present author favours  the causal interpretation and its extension to boson fields, both of  which provide a description of the physical mechanisms leading to the final experimental result in a way that is  fully consistent with the mathematical formalism of the quantum theory. It is not our purpose here  to discuss the causal interpretation or its extension in detail, so we will restrict ourselves to some brief comments. The causal interpretation provides a model for the tunneling of particles (impossible classically) in terms of the quantum potential \footnote{Though the $R$-field gives rise to the quantum potential, the $S$-field can also explain tunneling since the $R$ and $S$-fields codetermine one another. See ref. \cite{DEWD} for a computer model of quantum tunneling based on the causal interpretation.}.  The model is therefore in terms of both wave and particle concepts simultaneously, so it may be thought that GHA experiment provides evidence for the causal interpretation. But again, this is not so, since the wave behaviour is never actually measured by an intermediate measurement. A more accurate model of the electromagnetic field is provided by the extension of causal interpretation to the electromagnetic field. In this model photons (indeed, all bosons) are fields, never particles. In this case, a single photon  divides at the first prism face and tunneling is modeled in much the same way as in classical wave models (though there are still R and S-fields which guide the behaviour of the electromagnetic field, but these are now functions of the normal mode coordinates of the field). The entire GHA experiment is thus described purely in terms of a wave model. Perfect anticoincidence results, not because a photon followed a single path, since it split and followed both paths, but because of the nonlocal absorption   of the photon by an atom in one or other photomultiplier \cite{K06, K85, BHK87, K94, PNK2005}. We conclude that in the absence of an intermediate measurement in addition to the final measured result (anticoincidence), the GRA experiment neither refutes BPC, nor confirms one or other of the causal models, i.e., the experiment cannot  differentiate between the different interpretations.

\section{The Brida, Genovese,  Gramegna, and  Predazzi Experiment}

Some authors \cite{Ghose1999, UM96} questioned the statistical accuracy of the results of the Mizobuchi and Ohtak\'{e} experiment. This prompted Brida et al to carry out an improved experimental test \cite{BGGP04} addressing the criticisms of the Mizobuchi and Ohtak\'{e} experiment. In the BGGP experiment,  birefringence replaces tunneling through a gap between two prisms as the phenomenon that in classical theory can only be described by the wave concept. The  single photon source uses type I parametric fluorescent light generated by  a UV pump laser which is then passed through a nonlinear crystal. It is arranged so that only one photon at a time passes through a birefringent calcite crystal. Through the crystal the photon path splits along the ordinary or extraordinary directions. After exiting the crystal the two possible paths are directed to two different photodetectors. The perfect anticoincidence predicted by quantum mechanics was again observed. The perfect anticoincidence is consistent with a particle picture. Brida et al inferred wave behaviour from the fact that the photon underwent the birefringent phenomenon that classically can only be explained by wave behaviour. Brida et al, like GHA, claimed that both wave and particle behaviour is observed in the same experiment thus refuting complementarity.

But, the Brida et al variation of the GHA experiment suffers from the same criticism as the GHA experiment. The particle behaviour is consistent with the observed experimental results. The wave behaviour is not  measured, but only  inferred from the fact that in classical theory birefringence can only be explained by a wave theory. But, as for the GHA experiment, the description in terms of BPC is neither constrained by classical theory nor requires any other mechanism for birefringence, especially since classical concepts are not attributed physical reality. BPC  only requires that the classical concept  be consistent with the experimental arrangement and the final measured experimental result. There is no conceptual inconsistency in imagining a photon particle following either the ordinary or extraordinary direction. Again, the detection  in the photodetectors, consistent with particle behaviour according to the Bohr paradigm, fulfills the conditions for a measurement as defined in Sect.  \ref{WISAM}. On the other hand, the splitting of paths in the calcite crystal is nowhere amplified irreversibly and recorded on a macroscopic apparatus, nor is there a possibility of such a recording. Once again, wave behaviour is inferred, not measured. We see that the BGGP experiment is consistent with (defines) one and only one complementary concept; the particle concept. I again conclude that the  BGGP experiment is perfectly consistent with BPC.

As for GHA, two alternative descriptions of the BGGP experiment can be given in terms of the causal interpretation. In the nonrelativistic interpretation the quantum potential (equivalently the $S$-field) produces the splitting of paths along the  ordinary or extraordinary directions, with particles following one or other alternative path depending on their (hidden) initial position. The explanation of anticoincidence follows since the particle registers in one or other of the photodetectors. A truer model in terms of CIEM explains birefringence, as in classical physics, by a wave model (but involving the nonclassical $R$- and $S$-fields), and explains perfect anticoincidence by the nonlocal absorption of a photon.
 
\section{The Afshar Experiment\label{AFE}}

The aim in the Afshar experiment is to obtain undiminished image quality of two pinholes of a standard interference experiment  when  a wire grid is placed at the previously measured positions of the dark fringes of an interference pattern. Afshar argued that with the wire grid in place, and if interference occurs, there should be little or no loss of radiant flux, so that  the image quality should be comparable to the  image quality without the wire grid in place. In this way, Afshar devised a nondemolition measurement of interference fringes. The experiment indeed confirmed that there was almost no loss of resolution or total radiant flux. 

If interference did not occur, the wire grid would block some radiant  flux with a corresponding reduction of image quality and resolution. Since there was almost no loss of resolution or total radiant flux,  Afshar concluded that interference occurred prior to image formation. He attributed  the interference to wave behaviour.  Following Wheeler \cite{WHR78}, Afshar assumed that the image of a pinhole must be formed by photons coming from that pinhole, and considered this to constitute  a determination of the photon path. With this he associated particle behaviour. Afshar concluded that his experiment demonstrates the measurement of wave and particle behaviour in one and the same experiment in direct contradiction of BPC (and the duality relation).

\begin{figure}[h]
\unitlength=1in
\hspace*{1in}\includegraphics[width=2.8in,height=1.8in]  {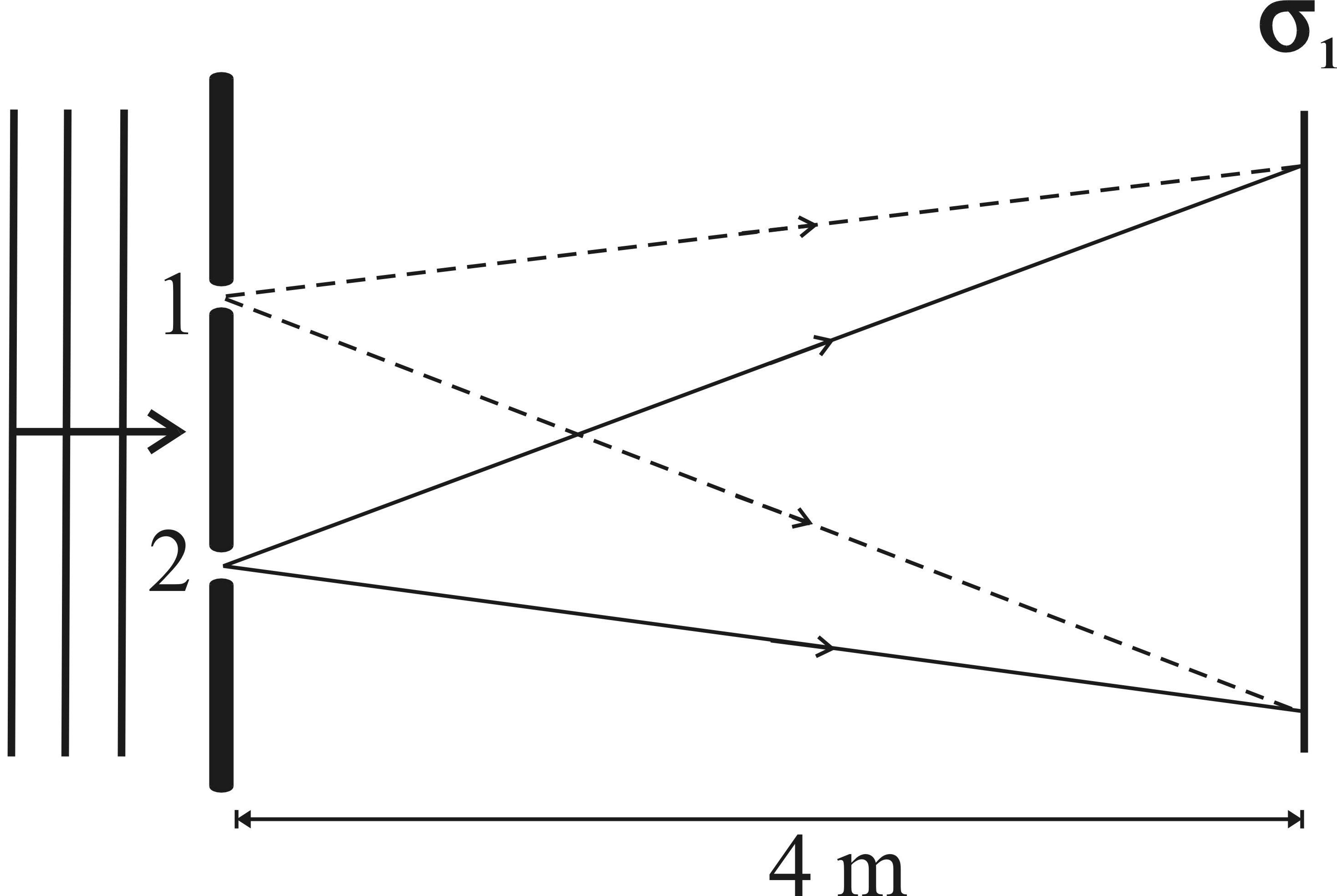}  
\caption{Afshar's experiment 1: Detection of interference by the photoplate at position $\sigma_1$ and the accurate measurement of the position of the dark fringes.}
\label{fig2}
\end{figure}

Afshar's experiment is split into three separate experiments. Experiment 1, shown in figure \ref{fig2}, is a standard two-pinhole interference experiment. Coherent and highly stable laser light of wavelength $\lambda=650$ nm is directed at a screen with two pinholes with diameters  $b=250$ $\mu$m, and separated by a distance $a=2000$ $\mu$m. The interference pattern is observed at a photoplate placed in the plane $\sigma_1$  4 m from the pinholes. The nonstandard feature of this experiment is the very accurate  measurement of the positions of the dark fringes of the interference pattern.

\begin{figure}[h]
\unitlength=1in
\hspace*{1in}\includegraphics[width=2.8in,height=1.8in]  {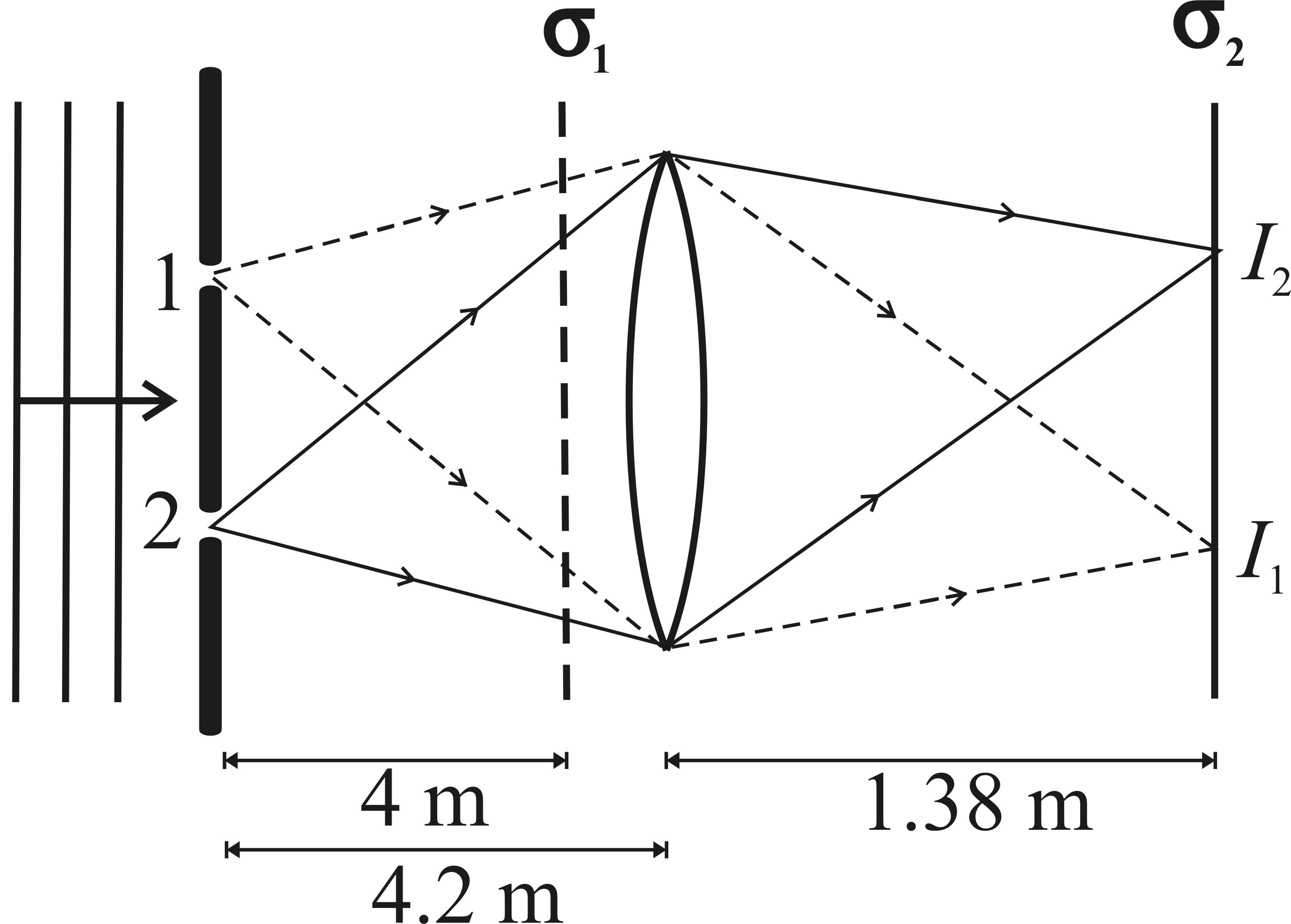} 
\caption{Afshar's experiment 2: The photoplate at $\sigma_1$ is removed and a lens is used to form images of the two  pinholes in the image plane $\sigma_2$.}
\label{fig3}
\end{figure}

In experiment 2, shown in figure \ref{fig3}, the photoplate  at $\sigma_1$ is removed and a lens of diameter 3 cm and focal length $f=100$ m is placed 4.2 m from the pinholes. The lens forms images of the two pinholes in the plane $\sigma_2$ 1.38 m from the lens. Experiment 2 acts as the control experiment with which the images formed with the wire grid in place are compared.

\begin{figure}[h]
\unitlength=1in
\hspace*{1in}\includegraphics[width=2.8in,height=1.8in]  {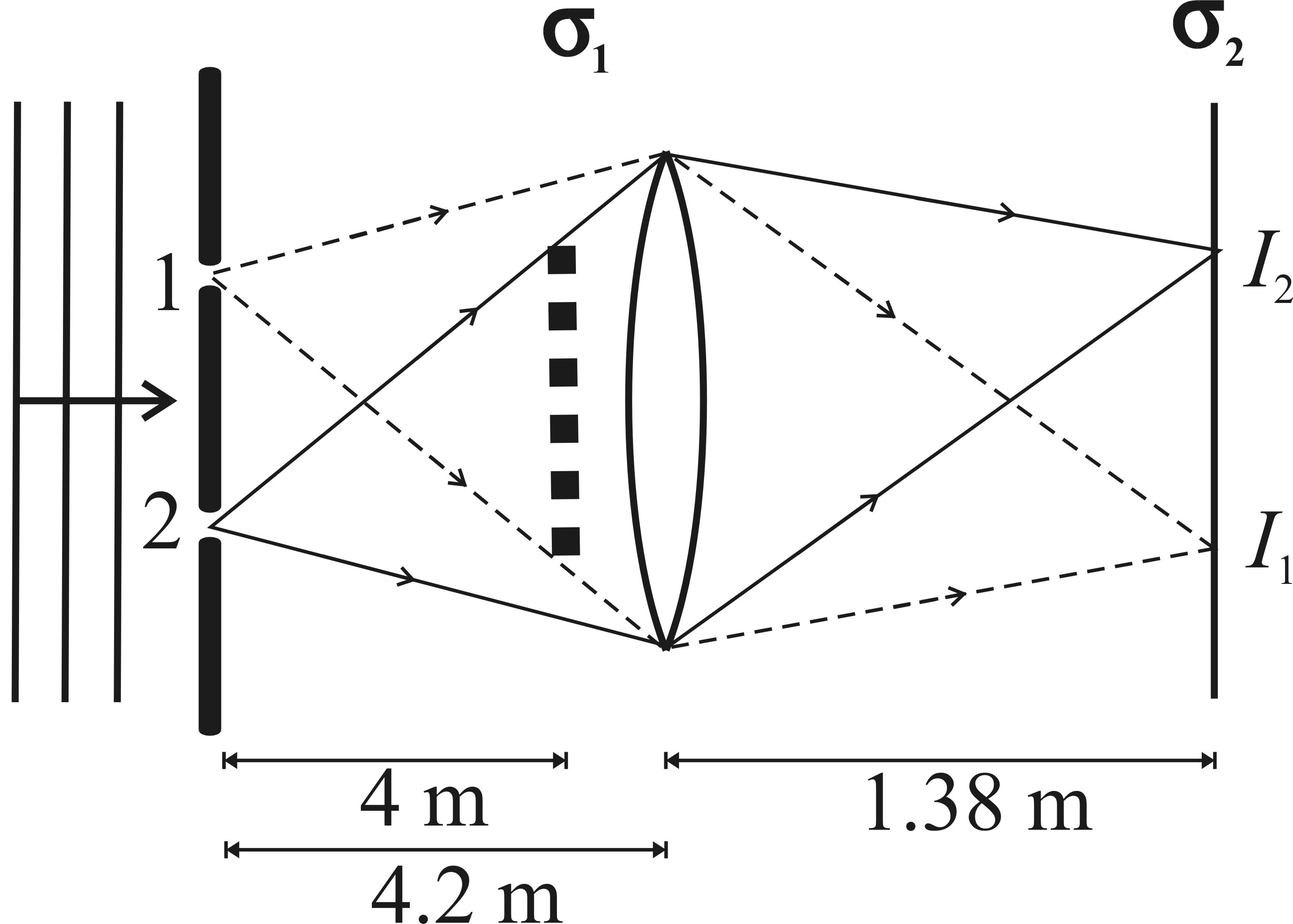}  
\caption{Afshar's experiment 3: A grid consisting of six wires is placed in plane $\sigma_1$ so that the wires coincide with the six central dark fringes of the interference pattern observed  in experiment 1. Almost no reduction in resolution or total radiant flux of the  images of the two pinholes formed in plane $\sigma_2$  was observed.}
\label{fig4}
\end{figure}

In experiment 3, shown in figure \ref{fig4}, a wire grid consisting of six wires is placed in the plane $\sigma_1$. The grid is positioned  so that the wires of the grid coincide with the previously measured positions of the dark fringes. Again, images of the two pinholes are formed at the image plane  $\sigma_2$. As I mentioned earlier, Afshar observed almost no loss in resolution or of total radiant flux as compared to the pinhole images obtained in experiment 2. He took this to be a definite detection of an interference pattern. To further justify his reasoning he closed pinhole 1, and obtained an image of pinhole 2 but with reduced intensity (i.e. less radiant flux formed the image compared to the control experiment). This showed two things. First, the image gives perfect path information so that $D=1$ and supports the interpretation, according to Afshar, that even with both pinholes open the images give perfect path information. Second, the reduction in radiant flux shows that there is no interference, so the wire grid scatters photons significantly. Since image quality (very few photons scattered) is restored when both pinholes are opened, Afshar considered this as further evidence that an interference pattern is detected with both pinholes open.

\begin{figure}[h]
\unitlength=1in
\hspace*{1in}\includegraphics[width=2.8in,height=1.8in] {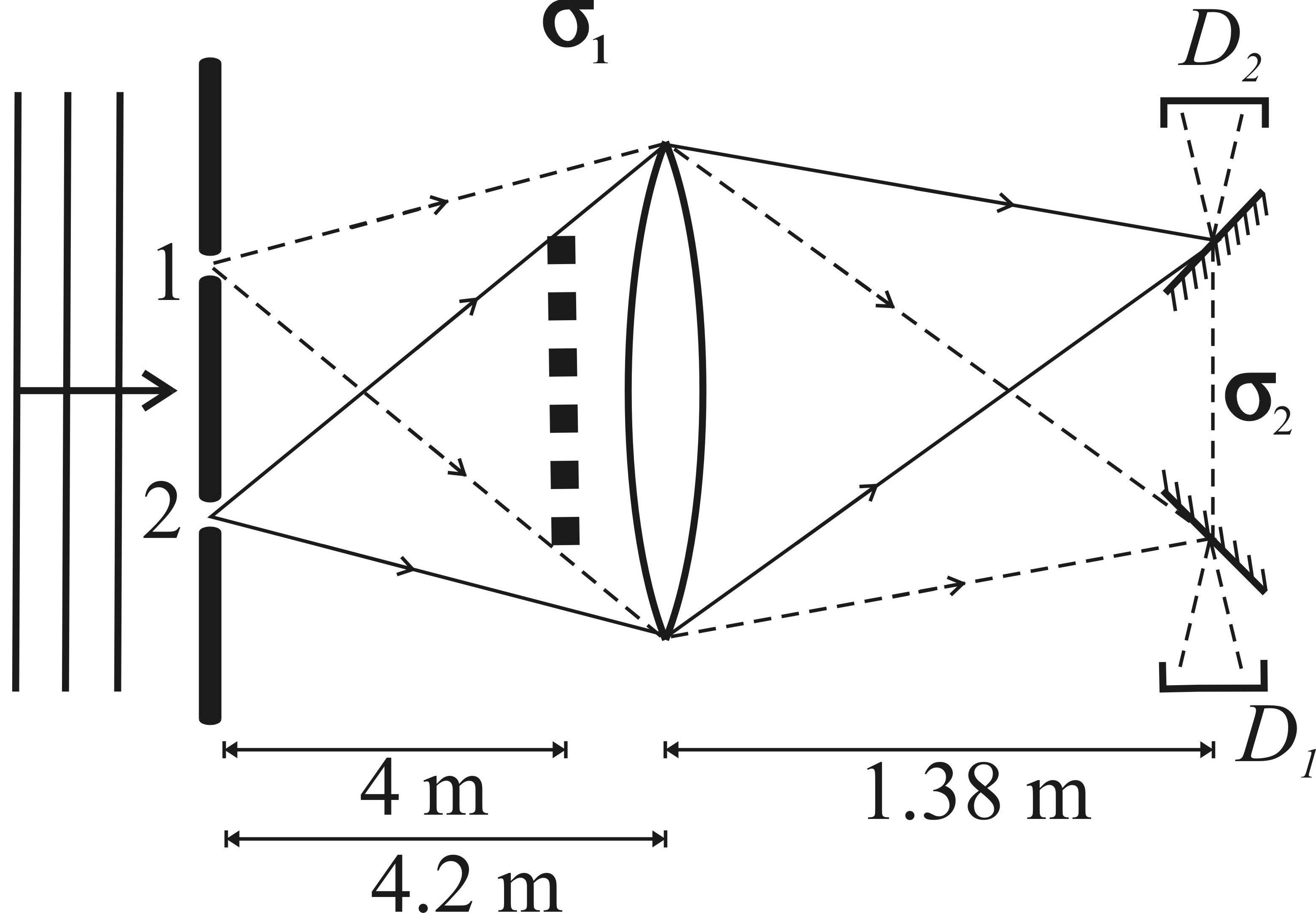}
\caption{Afshar's experiment 3.a: Placing either a photoplate  or two photodetectors directly in the plane  $\sigma_2$ leads to relatively high errors in the total radiant flux. Placing mirrors to direct the images  into high resolution photodetectors produces  more accurate results.}
\label{fig5}
\end{figure}

Placing high resolution photodetectors  in the positions shown in figure \ref{fig5}  produces more accurate measurements. 

In this way, Afshar concluded that the presence of the wire grid constitutes a (nondemolition) measurement of interference, and therefore wave behaviour. On the other hand, by assuming (not measuring) that a photon forming a pinhole image must have come from that pinhole, Afshar claims to have determined the  photon's path, and therefore to have observed particle behaviour. Thus, Afshar claims to have observed wave and particle behaviour in the same experiment in direct contradiction of BPC. 

Afshar equates the particle-wave duality relation to BPC, and measures  $D$ (or his version of the particle parameter) and $V$. He claims that his experiment  gives values  of both $D$ and  $V$ nearly equal to 1, so that $D^2+V^2$ has a value nearly equal to 2 in direct violation of the duality relation (\ref{PWDR}), and BPC.

Though I will argue below that the Afshar experiment fails to refute either BPC or the duality relation (\ref{PWDR}), it is nevertheless an interesting experiment from a technical perspective. The very interesting feature is the clear demonstration that interference takes place prior to  the formation of the image without any significant loss  of image quality. 

The experiment would by even more interesting if it could be repeated using a genuine single photon source of the type used in the  GRA experiment \cite{G86} and in the BGGP experiment  \cite{BGGP04}.   

\section{Critique of the Afshar Experiment\label{AFECRT}}
\label{CAE}
The Afshar experiment fails to refute either BPC or the duality relation for a similar reason that the GHA and BGGP experiments fail, namely, because one complementary concept (the wave concept) is measured while the other (the  particle concept) is inferred from an unjustified assumption. In the case of the Afshar experiment, the unjustified assumption is that a  photon forming a pinhole image must have come from that pinhole. Henceforth, I will refer to this as the Wheeler-Afshar Pinhole Assumption (WAPA). The WAPA is first of all not justified by the mathematical formalism of the quantum theory since the initial wave function is a superposition of wave functions emerging from each pinhole.  We recall Dirac's famous statement,  ``The new theory, which connects the wave function with probabilities for one photon, gets over the difficulty by making each photon go partly into each of the components. Each photon then interferes only with itself. Interference between two different photons never occurs"  \cite{DIR1958} pp. 9. A further argument against WAPA  is the description of the experiment according to the causal interpretation of boson fields applied to the electromagnetic field (CIEM) \cite{K06, K94, PNK2005}. According to this interpretation, which is a direct interpretation of the second quantized Maxwell equations, the photon, as I said earlier, is a field and so passes through both pinholes in a two pinhole arrangement. Further, the splitting of the photon at two pinholes, as deduced by Dirac and as required by CIEM, is indicated in the experiment of Tan, Walls and Collet in which homodyne detectors detect a single photon in both paths produced by a beam splitter \cite{TWC91}. A photon reaching the photodetector at $\sigma_2$ is spread over the images of the two pinholes, but registers as a ``spot'' in the photodetector. In CIEM, this is explained by the nonlocal absorption of the photon by atoms/molecules in the detector. The probability of absorption is highest at positions of highest intensity. When enough photon ``spots'' are detected, the images of the two pinholes emerge.

Perhaps  WAPA is suggested by the rays of geometric optics. However, the identification of photon trajectories with rays of geometric optics is arbitrary and has no theoretical justification. Again, the existence of CIEM emphasizes that such an association is erroneous, since, in this model, a photon is a field and passes through both pinholes.

Image formation and interference are  perfectly consistent with a pure wave concept, so that according to BPC the entire Afshar experiment is perfectly consistent with one and only one complementary concept; the wave concept. Thus, once we recognize that the claimed detection of path is based entirely on an arbitrary, unjustified  assumption, we see that the Afshar experiment is perfectly consistent with BPC.

Again, to complete the refutation  we need to justify what constitutes a measurement, as defined in Sect. \ref{WISAM}. Image formation at the photoplate clearly fulfills the conditions of a measurement. Though it is true that the pinhole images are formed by a large number of individual ``spot'' detections, image formation in this way is consistent with wave behaviour according to Bohr's paradigm. Even in the ordinary two-slit experiment the interference pattern is formed by a large number of photon by photon spot detections. The Bohr paradigm requires that the path of a photon leading to the spot detection be determined for the detection to be consistent with particle behaviour.

Further, the presence of the grid fulfills the conditions for a measurement, in this case, a null or nondemolition measurement. Why the presence of the grid constitutes a measurement of wave behaviour, while transmission through the gap or splitting of the path by a calcite crystal do not, is because there is a possibility for an irreversible, amplified detection to be recorded on a macroscopic apparatus (the wire grid). If interference did not take place photons would be absorbed   by the wire grid. Actually, if absorption did take place in would be revealed by two measurements. First, the absorption would be measured by a  reduction in the  intensity of the images. Second, the photon absorption, if it occurred, could be directly observed (measured) by looking at the wire grid after the experiment is concluded, since the absorbed photons would leave an irreversible amplified record on the macroscopic wire grid (e.g. by coating the grid with a photosensitive material). The absence of absorbed photons therefore constitutes a null or nondemolition measurement of interference.

Of course, there have been previous articles criticising Afshar's refutation. I mention a few such articles that are particularly notable: Kastner \cite{KAS05,KAS09}, Steuernagel \cite{STEUR}, Qureshi \cite{QUER} and Drezet \cite{DREZ}. These authors, like Afshar, equate BPC with the duality relation (\ref{PWDR}), and defend BPC by defending the duality relation. Inevitably these articles, including the present article,  have arguments in common, but each includes some original points or arguments from a different perspective. Taken together, these articles, in my view, constitute a very robust refutation that  Afshar's experiment contravenes either BPC or the particle-wave duality relation.

Flores and others have agreed with Afshar, and have worked on modified versions of Afshar's experiment, either independently \cite{FLOR09} or together with Afshar \cite{AFMK}. The modified experiments suffer from the same serious objections that Afshar's experiment suffers from.  

Drezet, in  his 2011 article \cite{DREZ} presents a detailed argument from the perspective of the duality relation to show that Afshar's experiment does not refute the duality relation, and  therefore also concludes  that BPC is not refuted by Afshar's experiment. I note this article in particular because Drezet discusses the relation of BPC with the duality relation and also the implications of BPC for the reality of trajectories from a less rigid perspective than I have done in the present article. Further, Drezet's arguments against the Afshar refutation of BPC have some common elements with my own.  One example, aside from the earlier point in connection with the duality relation (see footnote \ref{DREZFT}),  is that Drezet also emphasises that a photon forming a pinhole image need not originate from that pinhole. Also, in common is our conclusion that Afshar's experiment does not provide path information.

Kastner agrees with Afshar that a photon detection constitutes a path measurement, but disagrees with Afshar in that  he considers the state before the photon detection  as  a superposition state containing no path information. Kastner considers the photon detection in one image or the other as changing the photon's state from the superposition state containing no path information, to a ``path'' eigenstate. In this case, unlike Afshar, Kastner concludes that the ``path'' eigenstate  formed after the photon is detected does not reveal from which pinhole the photon came. Kastner thus concludes that $D$ in the Afshar experiment is zero. With $D=0$ and accepting Afshar's measurement of $V$ close to 1, Kastner concludes that the duality relation is preserved. Kastner's conclusion is in line with mine.  My only concern is Kastner's characterisation of the final photon state after detection as a``path'' eigenstate. Since, as Kastner himself states, the final photon detection does not reveal from which pinhole the photon came, in what sense is the detection a path eigenstate? If Kastner means by a path eigenstate, a position measurement of the photon, then, I would agree with him. But, since this detection does not reveal the photon path before detection, this detection cannot be viewed as a path eigenstate. The latter point is further emphasised, as Kastner himself correctly states, by the fact that prior to detection the photon is in a superposition state containing no path information.

Steuernagel criticizes Afshar's result $D^2+V^2=2$ based on a classical calculation (with results interpreted according to quantum theory). He saves the duality relation by calculating a value of $V$ to be close to zero  and by calculating a value of $D$ close to one, values that  are opposite to those of almost all other authors, including us. Aside from Steuernagel's calculation of $V$ close to zero being a direct contradiction of the actual measured  value in Afshar's experiment, a measurement considered genuine by  almost every other author, his calculation has been severely criticized by Kastner \cite{KAS09} and Flores \cite{FLOR08}. Further, implicit in  Steuernagel's analysis is the identification of photon trajectories with the rays of geometric optics.  This is the same identification that seems to underpin Wheeler's and Afshar's WAPA, mentioned and criticized above. Steuernagel's (and Afshar's) value for $D$ close to one based on the latter arbitrary, unjustified assumption cannot therefore be maintained. We conclude that Steuernagel reaches the correct conclusion, i.e., that Afshar's experiment does not refute the duality relation,  but for the wrong reasons.  Qureshi also concludes that Afshar's analysis is flawed. He saves the duality relation by arguing that the occurrence of the interference destroys the path information. This argument can be criticized, both from our arguments above, but also from the fact that a superposition state is required for interference to occur. Such a state does not carry path information. In other words, there was no path information in the first place to destroy.

\section{Conclusion}

We have tried to make a clear distinction, first, between particle and wave complementary concepts, which classically are mutually exclusive concepts, and complementary concepts which classically are simultaneously definable (and measurable) canonically conjugate variables, and second, between Bohr's principle of complementarity and the Wootters and Zurek reformulation of complementarity. We have argued that the duality relation (with its various definitions), which evolved from WZPC, does not have the fundamental significance that it is normally attributed. But, because of its clear mathematical significance it is worthwhile to note that the duality relation has been given a conceptually consistent interpretation by Jaeger, Shimony and Vaidman, an interpretation also consistent with BPC. I have suggested that a better interpretation of the duality relations should be based on an ontological interpretation of the quantum theory, such as  the causal interpretation. 

Concerning my main task, I have argued that quantum optical experiments which purport to refute BPC fail. I have tried to show that the reason such experiments fail is because only one complementary concept is actually consistent (defined) with the experimental arrangement and the measured result, while the other complementary concept is inferred based on an arbitrary and unjustified assumption or assumptions..

Although, I have argued against experimental refutations of BPC, I nevertheless feel that such tests are important, because, ultimately, such tests also test the quantum theory. Moreover, such tests seem to continually push technological limits. For both of these reasons such experimental tests are to be encouraged.

Finally, we have suggested that although BPC has not been experimentally refuted, it can be severely criticized on theoretical and conceptual grounds.  



%
%




\bibliographystyle{aps-nameyear}      

\end{document}